\documentclass[aip,amsmath,amssymb]{revtex4-1}
\usepackage{graphicx,dcolumn,bm,mathptmx,etoolbox}
\usepackage[utf8]{inputenc}
\usepackage[T1]{fontenc}

\makeatletter
\def\@email#1#2{%
 \endgroup
 \patchcmd{\titleblock@produce}
  {\frontmatter@RRAPformat}
  {\frontmatter@RRAPformat{\produce@RRAP{*#1\href{mailto:#2}{#2}}}\frontmatter@RRAPformat}
  {}{}
}%
\makeatother

\begin{document}

\title{Designing, Synthesizing and Modeling Active Fluids}
\author{Ilham Essafri$^{1,2}$, Bappa Ghosh$^{1,2}$, Caroline Desgranges$^2$ and Jerome Delhommelle}
\affiliation{Department of Biomedical Engineering, University of North Dakota, 243 Centennial Dr Stop 8155, Grand Forks, ND 58202, USA}
\affiliation{MetaSimulation of Nonequilibrium Processes (MSNEP), Tech Accelerator, University of North Dakota, Suite 2300, 4201 James Ray Drive, Grand Forks, ND 58202, USA}
\affiliation{Department of Chemistry, University of North Dakota, 151 Cornell Street Stop 9024, Grand Forks, ND 58202, USA}
\affiliation{School of Electrical Engineering and Computer Science, University of North Dakota 243 Centennial Dr Stop 7165, Grand Forks, ND 58202, USA}

\date{\today}

\begin{abstract}
We review recent advances in the design, synthesis, and modeling of active fluids. Active fluids have been at the center of many technological innovations and theoretical advances over the past two decades. Research on this new class of fluids has been inspired by the fascinating and remarkably efficient strategies that biological systems employ, leading to the development of biomimetic nano- and micro-machines and -swimmers. The review encompasses active fluids on both the nano- and micro-scale. We start with examples of biological active systems before we discuss how experimentalists leverage novel propulsion mechanisms to power nano- and micro-machines. We then examine how the study of these far-from-equilibrium systems has prompted the development of new simulation methods and theoretical models in nonquilibrium physics to account for their mechanical, thermodynamic and emergent properties. Recent advances in the field have paved the way for the design, synthesis, and modeling of autonomous systems at the nano- and micro-scale and open the door to the development of soft matter robotics.
\end{abstract}

\maketitle

\section{Introduction}

Fluids at low Reynolds numbers, {\it i.e.}, fluids for which viscous forces dominate, have gained enormous interest in the last several decades. There is indeed a whole world of living organisms that thrive under these conditions. A great example is {\it E. Coli}~\cite{berg1975bacteria, berg1975chemotaxis,berg1977physics,berg2003rotary}. Even if such fluids are well known to be important for engineers, and, in particular, for fluidized beds, their significance has been steadily increasing since the seventies. This is primarily due to the realization that, in such fluids, the force $\frac{\eta^2}{\rho}$, in which $\rho$ denotes the fluid density and $\eta$ the fluid viscosity, is independent from the inertial properties of the immersed system. As discussed by Purcell~\cite{purcell1977life}, this force will be able to tow anything, large or small, in fluids with a low Reynolds number. At the same time, moving in such viscous fluid requires some ingeniousness. Following the Purcell's 'scallop theorem’, if a low-Reynolds number swimmer executes a geometrically reciprocal motion, that is a sequence of shape changes that are identical when reversed, then the net displacement of the swimmer must be zero in an incompressible, Newtonian, fluid~\cite{lauga2011life,qiu2014swimming}. To quote Purcell, ``Fast, or slow, it exactly retraces its trajectory, and it’s back where it started''~\cite{purcell1977life}. Indeed, to be able to swim, bacteria use propulsion mechanisms, either cilia or flagella which beat or rotate, to make small moves, but also periodic deformations of their body to execute non-reciprocal motion and keep moving~\cite{purcell1977life, fischer2011magnetically}. Understanding the principles underlying such propulsion mechanisms and navigation strategies is still an outstanding challenge, with potential applications in medicine, among others~\cite{gompper2021motile}.

This observation has also opened the door to the development of a new field that is now known as active matter~\cite{ramaswamy2017active,marchetti2013hydrodynamics}. Active matter relies on the transduction of energy, often starting with the conversion of chemical energy, or ``fuel'', into mechanical energy, leading to the motion of the particles. This new bio-inspired research area leverages recent advances in self-propelled particles synthesis. Such particles can serve as elementary building blocks for active assemblies that mimic the response of groups, clusters, or colonies of biological swimmers. They, therefore, provide a path to study the onset of collective behavior and emergence in living systems. Experimental protocols have led recently to the design and synthesis of autonomous nanomachines and micromachines. They use different propulsion mechanisms such as phoresis, diffusiophoresis, or thermophoresis. These synthetic machines also pave the way for multifunctional materials and devices. They react and adapt to environmental cues and signals emitted from other synthetic machines, leading to the novel intelligent active materials design~\cite{aguilar2016review}. Achieving this goal will require a collective effort from multi-disciplinary team of scientists from biology, chemistry, physics, engineering, and mathematics to understand and control active matter.

Synthetic swimmers behave similarly to microorganisms as they change their swimming direction at regular intervals and interact with solid surfaces, as well as each other~\cite{dauchot2019chemical}. From a theoretical perspective, the theory of Brownian motion is perhaps the simplest approximation to model the motion of a small particle immersed in a fluid. While it appears to be random, it is possible to describe its motion using Langevin dynamics. The idea is to partition the total force acting on the Brownian particle with its environment (or heat bath) into a systematic part (or friction) and a fluctuating part (or noise). Each force is related to the other by the fluctuation-dissipation theorem~\cite{mazur1998fluctuations}. This provides a relation between the strength of the random noise (or fluctuating force) and the magnitude of the friction (or dissipation). As discussed by Zwanzig~\cite{zwanzig2001nonequilibrium}, it characterizes the balance between friction, which tends to drive any system to a completely "dead" state, and noise, which tends to keep the system "alive". When Brownian particles~\cite{babivc2005colloids} are self-propelled, we add an "active" swimming term to the equations of motion. Experiments show that such systems exhibit a larger diffusivity than passive Brownian particles. For instance, over 10 minutes, a passive particle may diffuse over a region of 35 $\mu m^2$, whereas the corresponding active particle may explore a region greater than 1$mm^2$~\cite{moran2019microswimmers}. The "active" additional force drives the system into a far-from-equilibrium state~\cite{hanggi2009artificial,hauser2015statistical}. Indeed, swimmers use energy from their environment, that they convert into directed motion~\cite{bechinger2016active,fodor2018statistical}. This results in a constant energy flow into the system and a time-reversal symmetry breaking which, in turn, has led to tremendous recent developments in the field of nonequilibrium physics~\cite{cates2012diffusive}. Models have also become more refined by incorporating hydrodynamics interactions, as well as interactions with complex environments. Such interactions take place at different spatial and temporal scales, prompting the development of new unifying principles at the mesoscale and of novel analytical tools to characterize emergent behavior in active assemblies~\cite{de2016lattice, zhang2011lattice,caballero2020stealth,shankar2018hidden,borthne2020time}.

The mini-review is organized as follows. In the first part, we provide an account of active fluids at the nanoscale and then focus our attention on active fluids at the microscale in the second part. For each part, we start by presenting examples of real-world biological systems, that provide the inspiration for the design of biomimetic synthetic materials, that are capable of responding and adapting to environmental cues like their real-life counterparts. We also discuss the latest advancements in nonequilibrium physics, as well as the recent advances in theoretical models and simulation methods, that promote the understanding of active matter and the rationalization of the novel emergent behavior observed in active systems. We finally summarize the main conclusions in the last section and how recent progress in the field paves the way for the development of an autonomous soft matter robotics.

\section{Active fluids at the nanoscale}

\subsection{Real-life systems}

{\bf Biological nanomachines}\\

There are numerous examples of molecular machines in biological and living systems~\cite{gibbs2011catalytic, zhao2018powering, kang2009single, wang2015dna, you2012autonomous, liu2013switching, mccullagh2011dna}. They exist under a myriad of forms and exhibit exceptional functionalities~\cite{chowdhury2008molecular,feringa2007art,kassem2017artificial,kistemaker2021exploring,astumian2010thermodynamics,sheetz1984atp,karki1999cytoplasmic,hirokawa1998kinesin}. For instance, nanomachines can synthesize complex molecules in the cell~\cite{steitz2008structural}. They can also serve as pumps~\cite{qiu2020pumps} and generate concentration differences, or as motors to convert chemical energy into directed motion. Nanomachines often work cooperatively~\cite{badjic2005multivalency,mayer2009molecular} to achieve even more complicated tasks. For instance, during muscle contraction, myosin II motors act cooperatively by binding to the same single actin filament and
pulling it against an external load~\cite{wagoner2021evolution}. More generally, cooperativity allows them to become biological factories~\cite{ozin2005dream} capable of controlling what happens in the cell.\\

\begin{figure}[ht]
\centering
\includegraphics[width=10cm]{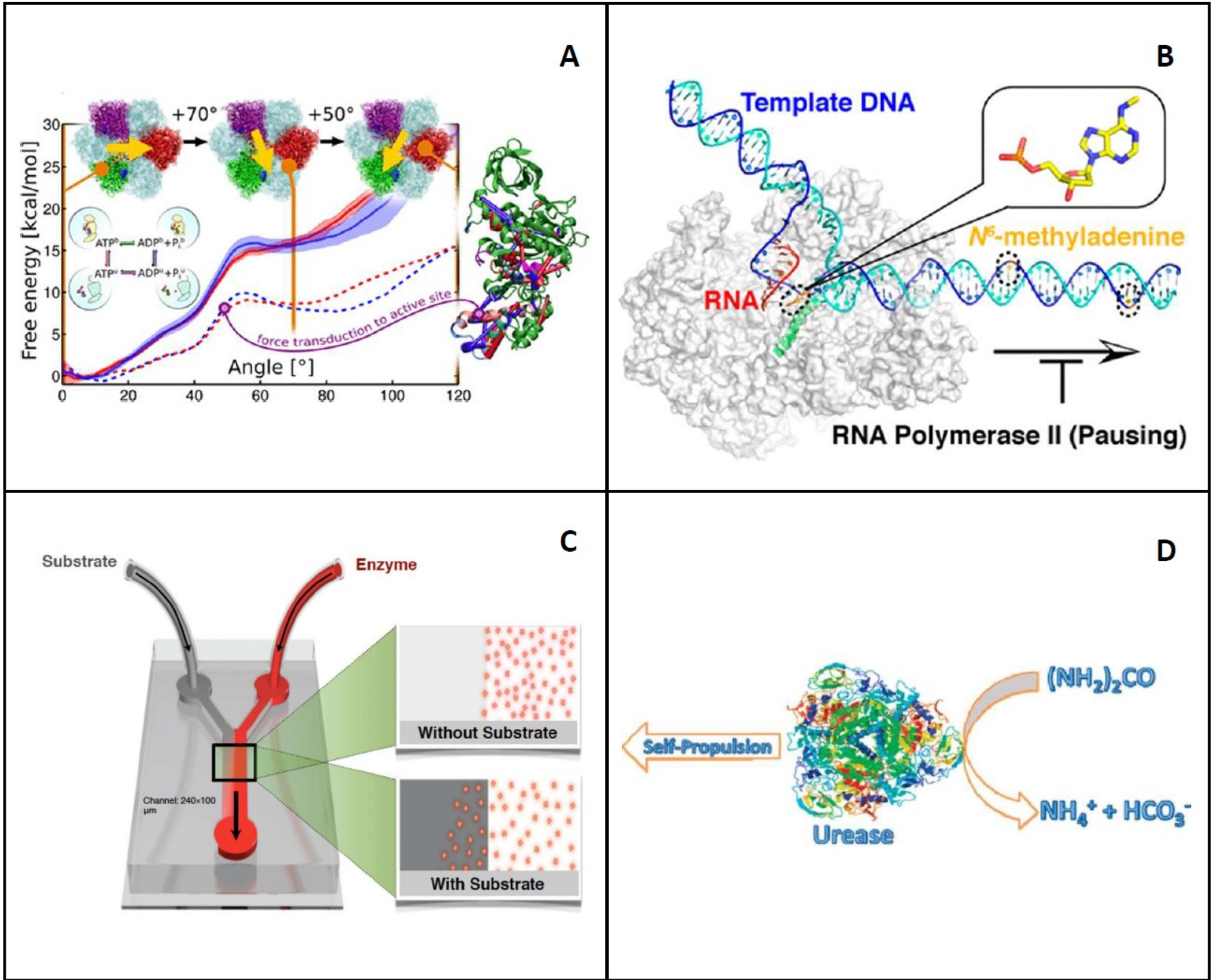}
\caption{Biological nanomachines. A) F$_1$-ATPase~\cite{czub2017mechanochemical}. The $\beta_{TP}$ subunit is shown in green and the $\beta_E$ subunit is shown in red. The orientation of the central asymmetric $\gamma$-shaft is shown with a yellow arrow in the two free energy minima (initial position at $0^\circ$ and metastable intermediate state at $70^\circ$) and at the end of the rotation cycle at $120^\circ$. Free energy profiles (solid lines) are shown in the presence of ADP (blue) or ATP (red) in the binding site of $\beta_{TP}$, and dashed lines show the corresponding profiles after the addition of the F$_o$-generated torque potential with a constant slope of 0.1~kcal$/^\circ$. Reprinted with permission from {\it J. Am. Chem. Soc.}~ {\bf 2017}, 139, 4025-4034. Copyright 2017 American Chemical Society. B) Impact of the N$^6$-Methyladenine (N$^6$-mA or 6 mA) epigenetic DNA modification of the DNA template on RNA polymerase II (pol II) transcription elongation, which causes site specific pol II pausing/stalling~\cite{wang2017epigenetic}. Reprinted with permission from {\it J. Am. Chem. Soc.}~ {\bf 2017}, 139, 14436-14442. Copyright 2017 American Chemical Society. C) Enzymes as Nanomotors: both catalase and urease enzymes move towards areas of higher substrate concentration generated by the Y-shaped microfluidic device, thereby exhibiting chemotaxis~\cite{sengupta2013enzyme}. Reprinted with permission from {\it J. Am. Chem. Soc.}~ {\bf 2013}, 135, 1406-1414. Copyright 2013 American Chemical Society.  D) Urease single-enzyme diffusion enhanced by substrate catalysis~\cite{muddana2010substrate}. Reprinted with permission from {\it J. Am. Chem. Soc.}~ {\bf 2010}, 132, 2110-2111. Copyright 2010 American Chemical Society.}
\label{fig1}
\end{figure}

{\bf Powering molecular machines with nanomotors}\\

One of the most fascinating enzymes in biology is the ATP synthase~\cite{boyer1997atp}. It is also the smallest known biological nanomotor~\cite{elston1998energy} and is found in almost all living organisms, including plants, animals, and bacteria. ATP synthase (see Fig.~\ref{fig1}A) is composed of two rotary motors, the proton-driven $F_0$ and the ATP-synthesizing $F_1$, that are coupled via elastic torque transmission\cite{junge2015atp}. This high-revving nanomotor mechanism utilizes the transport of protons to drive the synthesis of ATP. Another example of nanomachine is the organelle ribosome which consists of ribosomal RNA and protein~\cite{fischer2010ribosome}. Together they can read messenger RNAs and translate the encoded information into proteins. Nanomotors such as RNA polymerase~\cite{dehaseth1998rna} (see Fig.~\ref{fig1}B), DNA polymerase~\cite{saiki1988primer,kaguni2004dna}, and helicases~\cite{lohman1996mechanisms,caruthers2002helicase}, which deal with DNA and RNA reactions, walk along DNA strands to perform their functions. Molecular motors such as kinesins, myosins, or dyneins are responsible for directed motion. They use chemical energy to drive conformational changes, which lead to the active transport of material from one part of the cell to another. Dyneins~\cite{roberts2013functions}, fueled by ATP hydrolysis, generate force and movement along microtubules, allowing for the motions of cilia and flagella. Myosins~\cite{llinas2012myosin} are actin-based motor proteins that use cellular ATP to power interactions with actin filaments and create directed movements important in intracellular transport and cell division. Kinesins~\cite{hirokawa2009kinesin} are important molecular motors that directionally transport various cargos, including membranous organelles, protein complexes, and mRNAs. Another example is the ATPase active ion pump that can use some of the free energy released by the hydrolysis of ATP to pump sodium ions across the cell membrane. Proton-gradient-driven motors have also been observed to be the driving force behind flagellar filaments, which are used by many micro and nanoscale swimmers as propellers.~\cite{guo2016biological}\\

{\bf Enzymes as energy transducers}\\

A detailed understanding of how enzymes convert chemical energy into mechanical force has remained an outstanding challenge~\cite{arque2019intrinsic}. Enzymes act as catalysts in biological systems~\cite{wolfenden2001depth} (see Fig.~\ref{fig1}C). Once a substrate is bound to an enzyme's active site, the enzyme-catalyzed reaction is associated with a rapid turnover, as well as a high specificity and efficiency, that can power biological nanomachines~\cite{yang2016enzyme,patino2018fundamental,chen2019dual}. Recent work shows that enzymes could be at the center of the stochastic motion of the cytoplasm, the organization of metabolons, as well as the convective transport of fluid in cells~\cite{zhao2018powering}.  
Over the last decade, the development of enzyme-based nanomotors~\cite{sengupta2013enzyme} has been the focus of intense research. Indeed, recent studies have revealed the existence of free-swimming enzymes capable of moving in low Reynolds fluids. By harnessing chemical energy released through the enzymatic turnover of substrates, these free-swimming enzymes can generate enough mechanical force to power their motion and to enhance their diffusion~\cite{ma2016enzyme,riedel2015heat} (see Fig.~\ref{fig1}D). For instance, the diffusion of urease enzymes has been found to increase in the presence of urea~\cite{muddana2010substrate}. Further analysis shows that this increase depends on the concentration in the substrate and is weakened when urease is inhibited with pyrocatechol, thereby showing that the enhancement of diffusion results from the enzyme catalysis. After each turnover, the catalytic reactions generate an 'impulsive force', leading to a new kind of mechanobiological event. In addition, when subjected to a gradient in substrate concentration, enzymes move up, triggering chemotaxis at the molecular level~\cite{xu2021enzyme}. The design of "intelligent" enzyme-powered autonomous nanomotors, which have the ability to assemble and deliver cargo for biological applications, is thus now one of the most critical topics in nanotechnology. These discoveries could lead to the identification of the principles of fabrication and design of novel biomimetic molecular machines~\cite{zhao2018powering}.

\subsection{Synthetic systems}

{\bf Artificial biology}\\

\begin{figure}[ht]
\centering
\includegraphics[width=14.5cm]{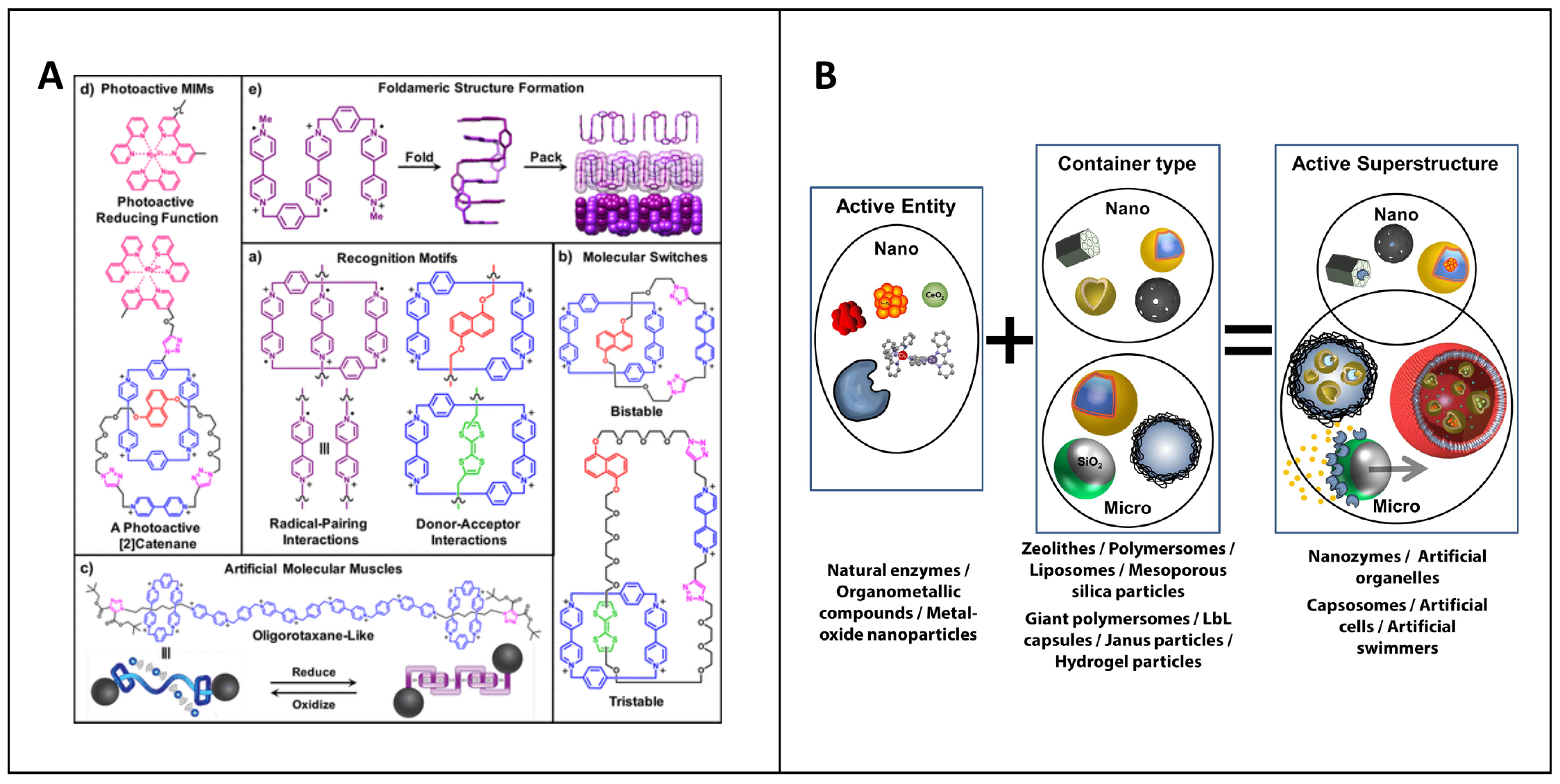}
\caption{A) Examples of multistimulus-responsive materials~\cite{wang2017introducing}. Reprinted with permission from {\it ACS Cent. Sci.}~ {\bf 2017}, 3, 927-935. Copyright 2017 American Chemical Society. B) Active entities are combined into superstructures to mimic therapeutic cells~\cite{itel2017enzymes}. Reprinted with permission from {\it Adv. Drug. Deliv. Rev.}~ {\bf 2017}, 118, 94-108. Copyright 2017 Elsevier.}
\label{fig2}
\end{figure}

Can we design and program nanosized synthetic machines to perform complex tasks similar to those performed by biological systems? The recent emergence of artificial biology~\cite{qian2021cell,wiester2011enzyme,tanner2011polymeric,zhang2008artificial} has focused on addressing this challenge. This new discipline aims at designing and engineering the structure and function of biological entities. This approach relies on a combination of biological and abiotic building blocks. For instance, it is now possible to integrate organelles, such as, {\it e.g.}, ribosomes) and biomolecules, such as, {\it e.g.}, enzymes, into an artificial membrane~\cite{stano2013semi, stano2014towards}. Another example is the recent realization of molecular walkers. Here, the idea is to start from proteins or nucleic acids. Chemical reactions are then used to drive conformational changes so that these molecular machines can walk on various materials ~\cite{leigh2014synthetic,shin2004synthetic,kay2007synthetic,hanggi2009artificial,antal2007molecular,leigh2014synthetic}. 
A related question is the following. Can we synthesize artificial molecular machines that can rival and potentially out-perform natural biological nanomachines? A possible route is supramolecular chemistry. The pioneering work by Sauvage, Stoddart, and Ferringa has revolutionalized the design and synthesis of molecular machines (see Fig.~\ref{fig2}A). Sauvage introduced a new type of bond, known as mechanical bond~\cite{bruns2016nature}, in chemistry. Stoddart developed mechanically interlocked molecules (MIMs) such as rotaxanes and catenanes~\cite{stoddart2017mechanically,sluysmans2018growing}, as well as molecular shuttle~\cite{anelli1991molecular} and molecular switches to control the motion of an artificial molecular motor~\cite{bissell1994chemically}. Feringa developed light-responsive rotors and used their rotary motions in mesoscopic and nanoscale applications, most notably in the well-known nanocar~\cite{feringa2017art}. Since then, more and more evolved molecular machines have been created, and the role played by response to an external stimulus, including redox conditions, pH, temperature and light has been emphasized~\cite{mavroidis2004molecular,ballardini2001artificial,balzani2006molecular,hoki2003molecular,hess2004powering,dreyfus2005microscopic,fletcher2005reversible,qu2005half,kay2007synthetic,balzani2008molecular,zambrano2009thermally}.  
Another route towards the design of synthetic nanomotors with directed motion is the construction of tiny chemically powered motors without moving parts. They rely on an asymmetric chemical reactivity, in which active particles harness chemical energy that they translate into work. Different types of nanomotors have been developed, such as nanowires~\cite{guo2018enhanced,wang2017precise}, helical motors~\cite{iamsaard2014conversion,wu2014turning,manesh2013nanomotor}, and nanorockets~\cite{gao2014synthetic,xu2020self,zha2018tubular}. To make up for the low Reynolds number surrounding them and counteract Brownian motion, nanomotors can be designed as fuel-dependent or fuel-independent nanomotors. The first category consists of nanomotors that catalytically turnover fuel from their environment to generate motion. They can act as motors or pumps and have demonstrated great significance in active transport at the nanoscale~\cite{chowdhury2008molecular, feringa2007art, kassem2017artificial, kistemaker2021exploring, astumian2010thermodynamics}. The other category of nanomotors extracts energy from external sources and converts it into motion.\\

{\bf Artificial enzymes: Nanozymes}\\

Recently, synthetic enzymes or nanozymes ~\cite{gao2007intrinsic} have emerged as exciting tools at the interface between chemistry, biology, materials, and nanotechnology. Such nanosized objects contain a part that mimics an enzyme activity that can act as a catalyst~\cite{itel2017enzymes}. Nowadays, there are more than $900$ nanomaterials classified as nanozymes. For instance, Fe$_3$O$_4$, CuO, Au, Ag, Pt, and Pd nanomaterials have been shown that exhibit a catalytic efficiency comparable to natural enzymes~\cite{yan2020nanozymology,chen2021control,singh2016cerium,gao2020kinetics}. Nanozymes are also major agents in nanorobotics and nanomedicine. Recent work has shown that they have the potential to allow for the building of the logic control, sensing, driving, and functioning system of nanorobots~\cite{jiang2018standardized}. For instance, catalase-like nanozymes are essential to artificial motility. They can catalyze the decomposition of H$_2$O$_2$ into molecular oxygen (O$_2$) and water (H$_2$O) and, as a result, power motion. Nanozymes can also help control motion. Recent work has shown that nanozymes decorated with targeting molecules can assist the motion and guidance of the nanomachine to the targeting position~\cite{xie2020prototypes}.\\

{\bf Fuel-dependent systems: translational and rotational motion}\\

Two general types of nanomotors can produce motion at the nanoscale. The first type encompasses nanomotors that operate thanks to a phenomenon known as self-electrophoresis. This concept was introduced by Mitchell in 1956~\cite{mitchell1972self} and the underlying idea is the following.  When a bacterium pumps ions across its membrane asymmetrically, it forms an electrical circuit. Indeed, if ions are pumped out at one end of the cell and back in at the other end, ions flow from the rear of the bacterium's body to the front and the organism moves forward. This self-generated flow field creates an autonomous motion which is from now on referred to as self-propulsion~\cite{mano2005bioelectrochemical, mirkovic2010fuel}. The first synthetic design of this type of autonomous nanomotor was first reported in 2004 by Paxton {\it et al.}~\cite{paxton2004catalytic}. They used a Pt-based nanomotor capable of powering a linear motion and showed that a platinum-gold nanorod underwent self-electrophoresis in a hydrogen peroxide solution. In other words, the Pt nanozyme, one of the metallic nanoparticles that possess catalase/peroxidase-like activity, catalyzes H$_2$O$_2$ to generate O$_2$ or oxidize other substrates. The concentration gradient in O$_2$ generates an interfacial tension gradient which, in turn, results in motion. Paxton {\it et al.} also reported the speed of the motor to be equal to approximately several body lengths per second.~\cite{paxton2006catalytically}.

Recent work has focused on designing novel catalytic nanomotors capable of different types of motion such as rotation~\cite{kim2016man}. Wang {\it et al.} used Tafel plots to predict the direction of motion of all possible bimetallic combinations through self-electrophoresis~\cite{wang2006bipolar}. Fournier-Bidoz {\it et al.} designed a self-powered synthetic nanorotor from barcoded gold–nickel nanorods, with the gold end anchored to the surface of a silicon wafer~\cite{fournier2005synthetic}. They observed circular movements at constant velocity as hydrogen peroxide fuel is catalytically decomposed into oxygen at the unattached nickel end of the nanorod. By varying the concentration of hydrogen peroxide and the length of the nickel segment, Fournier-Bidoz {\it et al.} controlled the angular velocity of the rotating nanorods. Moreover, when several hundred nanorods are present in the solution, the authors observed novel rotational behaviors. For instance, a nanorod rotating clockwise can undergo a collision with another nanorod and the resulting nanorod pair can rotate counterclockwise. New designs can also lead to novel mechanized functions. For instance, Solovev {\it et al.} reported autonomous and remotely guided catalytically self-propelled InGaAs/GaAs/(Cr)Pt nanotubes~\cite{solovev2012self}. These rolled-up tubes with diameters of 280-600 nm move in hydrogen peroxide solutions with speeds as high as 180 $\mu$ m s$^{-1}$. The effective transduction of chemical energy into translational motion allows these nanotubes to perform tasks such as cargo transport (see Fig.~\ref{fig2}B). Furthermore, while cylindrically rolled-up tubes move in a straight line, asymmetrically rolled-up tubes follow a corkscrew-like trajectory. These nanotubes can thus drill and embed themselves into biomaterials.\\

{\bf Fuel-free nanomotors}\\

Fuel-free nanomotors~\cite{xu2017fuel} have become leading candidates for applications in nanomedicine. Unlike fuel-dependent nanomotors, their propulsion mechanisms are biocompatible and sustainable. Instead of relying on a chemically-powered propulsion, fuel-free nanomotors leverage external stimuli such as, {\it e.g.}, magnetic, chemical, thermal, or electrical fields. Several groups have been exploring fuel-free nanomachine propulsion mechanisms, including the utilization of magnetic~\cite{zhang2009artificial, ghosh2009controlled,gao2010magnetically,tottori2012magnetic,gao2014bioinspired}, electrical~\cite{chang2007remotely,loget2010propulsion,calvo2010propulsion}, optical~\cite{liu2010light} and ultrasonic~\cite{wang2012autonomous,kagan2012acoustic, garcia2013functionalized} fields.
Magnetically-driven nanomotors are noteworthy devices since they require field strengths harmless to humans. They are particularly promising in a variety of {\it in vivo} biomedical applications. Ghosh {\it et al.} recently reported the first “voyage” in human blodd of magnetic nanomotors, based on conformal ferrite coatings~\cite{venugopalan2014conformal}. Other outstanding applications of magnetically-driven nanomotors deal with cellular internalization. Cellular functions and physiology are dependent on the rheological properties of the cell. Moreover, since the cellular interior is constantly reorganizing, this contributes to a highly complex mechanical environment characterized by heterogeneities across multiple length scales. Helical nanomotors recently helped monitor the motion of the cytoskeleton inside the cell. More specifically, researchers show that variations of the hydrodynamic pitch in the helical propulsion define different elastic relaxation time scales, corresponding to the locations and motions of the cytoskeleton~\cite{spagnolie2013locomotion}. In other studies~\cite{pal2018maneuverability}, helical nanomotors driven by small rotating magnetic fields allowed for the exploration of the interior of cancerous cells.

Ultrasound-driven (US) nanomotors belong to another class of synthetic fuel-free nanomotors. They are propelled through acoustic fields and are very effective in intracellular delivery. Indeed, US-powered nanomotors possess sufficient force for penetrating the cellular membranes, rapidly internalizing into cells, and actively delivering therapeutic cargoes. Mallouk’s group demonstrated the first effective internalization of gold nanorods motors into HeLa cells after 24 h incubation. It was also possible afterward to activate the intracellular propulsion of these internalized nanomotors with an acoustic field, involving axial propulsion and spinning~\cite{wang2014acoustic}. Over the past years, several studies have demonstrated the advantages of acoustic nanomotors for intracellular applications~\cite{esteban2016acoustically,hansen2018active}. For instance, US-propelled nanomotors deliver small interfering RNA (siRNA) payloads inside cells for gene-silencing applications. More recently, acoustic nanomotors have been designed to transport oxygen inside cells toward promising therapeutic applications~\cite{zhang2019nanomotor}. Additionally, the active internalization and motion of acoustic nanomotors inside cells have been exploited for enhanced intracellular sensing of disease biomarkers, including specific nucleic acids and proteins~\cite{lee2016biosensors,gao2014graphene}.

Finally, nanomotors propelled by light are very promising devices~\cite{xu2017light}. Light allows for the manipulation of nanomachines with spatial and temporal precision, as experimentalists can readily modulate light intensity, frequency, polarization, and propagation direction. This, in turn, enables excellent controllability and programmability of these nanomotors. Moreover, photo-catalysis can help design new light-powered nanomotors. The idea is to use already well-known photo-electrochemical reactions to advance the development of these new nanodevices. In other words, a light-powered nanomotor is motorized thanks to the photovoltaic effect. The electric current is then converted into propulsion through electrokinetic mechanisms~\cite{wang2018light}. Wang {\it et al.}~\cite{wang2017silicon} proposed a visible-/near-infrared light-driven nanomotor based on a single silicon nanowire. The silicon nanomotor harvests energy from light and propels itself by self-electrophoresis. Importantly, due to the optical resonance inside the silicon nanowire, the spectral response of the nanowire-based nanomotor can be modulated by the nanowire's diameter. This pioneering work provides new opportunities to develop novel functions such as multiple communication channels in nanorobotics and controllable self-assembly. Other examples based on light within the "therapeutic window" and thus compatible with living tissues have been developed recently. Nelson {\it et al.} used black TiO$_2$ and were able to design a visible-light-driven nanomotor~\cite{jang2017multiwavelength}. Tang {\it et al.} fabricated nanomotors with silicon, which can be driven by visible and near-infrared radiation at ultralow intensity. They recently designed a visible-light-propelled nanotree-based microswimmer~\cite{zheng2017orthogonal} using the principles of the dye-sensitized solar cell. By loading dyes, they were able to have complementary absorption spectra and thus control the navigation of the microswimmers.\\

{\bf Directed motion for therapeutics}\\

Here, we examine directed motion created by external fields to guide nanomotors. External stimuli can guide and direct nanomotors towards an area of interest with a high efficiency. Here, the idea is to control the motion of a group of particles rather than guiding every single one independently. This new approach offers exceptional opportunities in drug transport and delivery~\cite{gao2014synthetic}. Several methods for motion control, such as magnetic guidance, thermal control, chemical response, and phototaxis, have already been proved to have a significant impact.\\
Catalytically-powered nanomotors are starting to play a significant role in cargo towing. Indeed, magnetic guidance has shown tremendous progress on cargo-carrying catalytic nanomotors for different loading and unloading mechanisms. Kagan {\it et al.}~\cite{kagan2010rapid} presented the first example of nanoshuttles for the transport and release of drugs. This pioneering study illustrated that catalytic nanowire shuttles could readily pick up drug-loaded particles and transport them over predetermined routes toward target destinations. Moreover, the combination of carbon nanotubes with nanomotors, such as CNT-based nanowire motors~\cite{burdick2008synthetic}, showed excellent results when transporting ‘heavy’ therapeutic cargo. In this case, the nanomotors pick up, transport, and release varying-sized drug carriers towards predetermined destinations. Magnetic guidance also helped navigate and deliver a cargo with precision inside channel networks, with a drastic acceleration of the nanomotor ~\cite{laocharoensuk2008carbon,lu2019highly} and a speed close to natural biomolecular motors ($50-60~\mu$m/s). Calvo-Marzal {\it et al.} designed an electrochemically switch to control and fine-tune the speed of a catalytic nanomotor. Here, the potential-induced motion control is attributed primarily to changes in the local oxygen level in connection with the interfacial tension gradient. Such reversible voltage-driven motion represents an attractive approach for the on-demand regulation of artificial nanomotors and opens the door to new and exciting operations of these nanoscale devices~\cite{calvo2009electrochemically}.

\subsection{Theory and simulations}

{\bf Continuum theory for phoretic propulsion}\\

The propulsion of synthetic nanomotors relies on phoretic mechanisms~\cite{anderson1984diffusiophoresis,shaebani2020computational,moran2017phoretic}. For instance, a Janus particle can be described as a spherical particle with a catalytic, or reactive site, on its surface~\cite{golestanian2005propulsion}. Reactions take place at the catalytic site and result in concentration changes in the chemicals contained in the surrounding fluid. In turn, these concentration changes give rise to chemical gradients that trigger a hydrodynamic flow in the vicinity of the surface and lead to the potion of particles with a momentum of the same magnitude, and in the opposite direction, as that for the surrounding fluid because of the conservation of momentum. Such a motion is termed as self-diffusiophoresis~\cite{popescu2016self}. Several theories have been developed to account for phoretic mechanisms~\cite{eloul2020reactive}, starting with the pioneering contribution of Derjaguin and Sidorenkov~\cite{derjaguin1947kinetic}, including those triggered by gradients due to electric fields~\cite{illien2017fuelled,zemanek2019perspective} or temperature gradients~\cite{lattuada2019thermophoresis,burelbach2018unified} in addition to concentration gradients~\cite{stark2018artificial}.

Classical hydrodynamics can be invoked to derive the underlying equations, together with the boundary conditions, and determine the self-propulsion velocity of, {\it e.g.}, the Janus particles described above~\cite{golestanian2005propulsion,popescu2016self,julicher2009generic,popescu2010phoretic,reigh2015catalytic,gaspard2018fluctuating,shaebani2020computational}. As discussed by Slomka and Dunkel~\cite{slomka2015generalized}, one can write the swimmer velocity field ${\mathbf u}$ as a sum of two terms, ${\mathbf v}$ and $v_0{\mathbf P}$, in which ${\mathbf v}$ is the solvent velocity field, ${\mathbf P}$ is the local mean orientation of the swimmer and $v_0$ the self-propulsion velocity relative to the solvent flow. The dynamics of the solvent velocity field ${\mathbf v}$ is given by the Navier-Stokes equation (equations of conservation of momentum and mass) for an incompressible solvent flow as~\cite{hinch1988hydrodynamics,lauga2009hydrodynamics}
\begin{equation}
\begin{array}{ccc}
\rho (\partial_t + {\mathbf v}\cdot \nabla){\mathbf v} & = & -\nabla P + \eta \nabla^2 {\mathbf v}\\
\end{array}
\label{NS1}
\end{equation}
\begin{equation}
\begin{array}{ccc}
\nabla \cdot {\mathbf v} & = & 0\\
\end{array}
\label{NS2}
\end{equation}
in which $\rho$ denotes the mass density, $P$ the scalar pressure and $\eta$ is the viscosity. The stress tensor $\mathbf {\sigma}$ is given by ${\mathbf \sigma} = P {\mathbf 1} + \eta [ \nabla {\mathbf u} + (\nabla {\mathbf u})^T]$, in which ${\mathbf 1}$ is the identity tensor. The hydrodynamic force ${\mathbf F}$ and torque ${\mathbf L}$ acting on the swimmer are found by integrating
over the surface $S$ of the swimmer~\cite{lauga2009hydrodynamics}
\begin{equation}
\begin{array}{ccc}
{\mathbf F(t)}=\int \int_S {\mathbf \sigma} \cdot {\mathbf n}~dS
\end{array}
\end{equation}
\begin{equation}
\begin{array}{ccc}
{\mathbf L(t)}=\int \int_S  {\mathbf x} \times {({\mathbf \sigma} \cdot {\mathbf n})}~dS
\end{array}
\end{equation}
where ${\mathbf x}$ is the position on the surface $S$ and ${\mathbf n}$ the unit vector normal to S into the fluid.

Eqs.~\ref{NS1}~and~\ref{NS2} assume that the suspension has reached a quasi-equilibrium state. As discussed by Slomka and Dunkel~\cite{slomka2015generalized}, this state is such that the net momentum transfer between swimmers and the surrounding fluid is negligible, the active particles can be regarded as force-free and the solvent flow is driven by the stress field $\mathbf{\sigma}$ created by the swimmer~\cite{najafi2004simple,pooley2007hydrodynamic,drescher2011fluid,guasto2010oscillatory,lauga2009hydrodynamics}. We add that recent work has started to focus on the impact of an increase in the size of active particles on their dynamics change as inertial effects start to become more significant~\cite{redaelli2022unsteady}. In such cases, inertial effects due to, {\it e.g.}, the unsteady acceleration of a swimmer, are described by the unsteady time-dependent Stokes equation~\cite{wang2012unsteady}. 

\begin{figure}[ht]
\centering
\includegraphics[width=14.5cm]{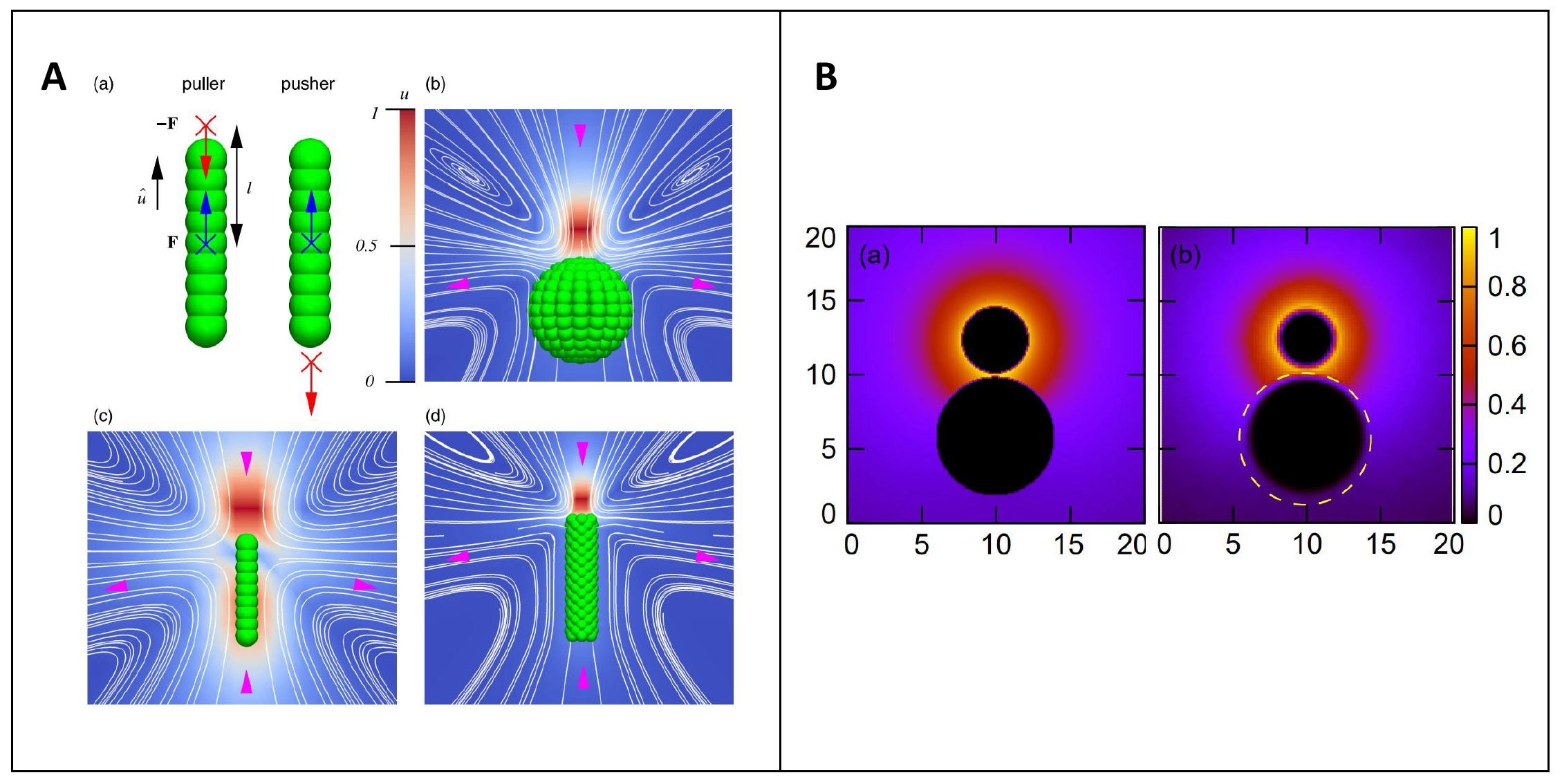}
\caption{A) Examples of "raspberry" swimmers, with in (top left panel) a sketch of the construction of pusher and puller raspberry swimmers (rods). A force ${\mathbf F}$ (blue arrow) is applied to the central bead (blue cross) in the direction of the symmetry axis $\hat u$ (black arrow). A counter-force${- \mathbf F}$ (red arrow) is applied to the fluid at a point $l \hat u$ (red cross), with $l$ the dipole length. $l>0$ corresponds to a puller and $l<0$ to a pusher. The other panels show the flow field around puller raspberry swimmers. Streamlines are shown in white and magenta arrow heads indicate the flow direction. Reprinted with permission from {\it J. Chem. Phys.}~ {\bf 2016}, 144, 134106. Copyright 2016 American Institute of Physics. B) Comparison between the normalized concentration field for the self-propulsion reaction product from (left) the continuum theory and (right) MD-MPCD simulations~\cite{reigh2015catalytic}. Reprinted with permission from {\it Soft Matter}~ {\bf 2015}, 11, 3149-3158. Copyright 2015 Royal Society of Chemistry.}
\label{fig3}
\end{figure}

The next step consists in introducing the Reynolds number $R_e=(VL)/\nu$), in which $\nu$ is the kinematic viscosity, $V$ and $L$ and characteristic length and speed scales in the system, and take the low Reynolds number regime limit ($R_e \to 0$) to obtain the Stokes equation from Eq.\ref{NS1}
\begin{equation}
\begin{array}{ccc}
\nabla P + \eta \nabla^2 {\mathbf v} & = & 0\\
\end{array}
\label{STK}
\end{equation}
\begin{equation}
\begin{array}{ccc}
\nabla \cdot {\mathbf v} & = & 0\\
\end{array}
\end{equation}
The flow field obtained from the solution to the Stokes equation, or Oosen tensor, is known as a Stokeslet, and gives rise to what is known as a dipole swimmer model, in which the nanomotor is approximated by two point forces with opposite directions. The flow field is then given by an expansion, in which dipole contributions provide the leading term far away from the self-propelled particle~\cite{elgeti2015physics,spagnolie2012hydrodynamics,berke2008hydrodynamic}. The resulting flow lines (see, for instance, the flow lines shown in Fig.~\ref{fig3}A) lead to the onset of effective interactions with surfaces and other nanomotors, which can be either attractive (inflow) or repulsive (outflow). Flow-induced interactions have been found to account for, among others, the nematic arrangement of elongated self-propelled particles~\cite{saintillan2007orientational,saintillan2008instabilities}.

The asymmetric catalytic reactions that occur at the surface of the self-propelled particles generate a mechanochemical coupling between the flow field and the concentration fields of the reactants. The concentration field of any given solute $i$ can be determined from a reaction-diffusion equation. This equation follows from the conservation equation for $i$, given by
\begin{equation}
\partial_t c_i = D \nabla^2 c_i + R(c_i)
\label{conc}
\end{equation}
in which $c_i$ is the concentration in solute $i$, $D$ the diffusion coefficient and $R(c_i)$ stands for the changes in concentration arising from the chemical reactions occurring at the catalytic side of the nanomotor. Strictly speaking, Eq.~\ref{conc} is valid in the low P\'{e}clet number ($P_e$ regime, or, in other words, that the ratio of the solute advection to diffusion, measured by $P_e=LV/D$, is very small~\cite{anderson1982motion}. Assuming that the concentration field has reached the stedy-state, Reigh and Kapral were able to determine the concentration field (see Fig.~\ref{fig3}B) and, from there, to determine the propulsion velocity for the nanomotor as a function of the concentration field and of the rate constant for the self-propulsion reaction~\cite{reigh2015catalytic,robertson2018synthetic}.

Gapsard and Kapral recently proposed a fluctuating chemohydrodynamics theory that accounts for the stochastic motion of self-diffusiophoretic particles~\cite{gaspard2018fluctuating,gaspard2019thermodynamics}. They derived equations of motion for the stochastic dynamics
and reaction of an active Janus particle self-propelled by diffusiophoresis. Specifically, using Green-function methods and the Faxen theorem~\cite{gaspard2018fluctuating}, they obtained the frequency-dependent force, torque, and reaction rate from the boundary conditions and the fluctuating chemohydrodynamic equations. This has led to the identification of coupled Langevin equations for the translation and rotation of self-propelled particles, as well as for the reaction. They showed that the equations so obtained are consistent with the Onsager-Casimir reciprocal relations between affinities and currents, thereby providing a thermodynamically consistent picture for self-diffusiophoretic particles.\\

{\bf Coarse-grained simulation methods for nanomotors}\\

Simulations provide a fascinating alternative to visualize, understand and rationalize the mechanisms underlying the operation of nanomotors and their collective behaviors. Furthermore, they can provide access to properties that are difficult to obtain in experiments. Golestanian {\it et al.} proposed one of the first models for nanomotors~\cite{golestanian2005propulsion,golestanian2007designing,howse2007self}. The nanomotor is modeled as a spherical particle, with a reactive patch on its surface, and its motion driven by the asymmetric distribution of reaction products. This success inspired the development of one of the most popular and simple nanomotor models, known as the sphere-dimer motor~\cite{ruckner2007chemically,valadares2010catalytic,reigh2015catalytic,tao2008design,novotny2020nanorobots,chen2021dynamics,chen2018chemically,chen2016chemotactic,tao2010swimming,zhou2020self,shi2018pair,colberg2017many,reigh2018diffusiophoretically,yu2018chemical}. The simulation of the sphere-dimer motors is then carried out using a hybrid method, that integrates the motion of the sphere-dimer particles with a classical molecular dynamics scheme~\cite{allen2017computer} and the time-dependent evolution of the surrounding solvent with a multiparticle collision dynamics scheme~\cite{malevanets1999mesoscopic,malevanets2000solute,malevanets2004mesoscopic,kapral2008multiparticle,ripoll2005dynamic,allahyarov2002mesoscopic,zottl2018simulating,gompper2009multi}. The sphere-dimer consists of two Lennard-Jones sites~\cite{allen2017computer}, labeled as {\it C} for the catalytic sphere and as {\it N} for the noncatalytic sphere, which are held together via a holonomic contraint at a fixed distance $d_{CN}$~\cite{colberg2014chemistry,robertson2018synthetic}. Each of the Lennard-Jones sites interact with other motor sites and solvent particles via the usual Weeks-Chandler-Andersen modification~\cite{allen2017computer} of the Lennard-Jones interaction potential
\begin{equation}
V_{sf}=4\epsilon_{sf} \left[ {\left({\frac{\sigma_sf}{r}} \right)^{12}} - {\left(\frac{\sigma_sf}{r} \right)^6} + {\frac{1}{4}} \right]    
\end{equation}
in which the index $s$ denotes either $C$ or $N$, $f$ refers to a fluid particle, and $\epsilon{sf}$ and $\sigma_{sf}$ are the Lennard-Jones parameters for the depth of the attractive well and for the exclusion diameter of the underlying Lennard-Jones potential. The minimal number of types of fluid particles is 2, corresponding to type $A$ before the fluid particle has undergone the irreversible chemical reaction $A \to B$ that accounts for the self-propulsion of the motor and to type $B$ after the reaction has taken place.\\

The fluid particles are coarse-grained into effective particles and the fluid-fluid interactions and chemical reactions that take place within the fluid are then implemented through a reactive multiparticle collision dynamics~\cite{kapral2008multiparticle,malevanets1999mesoscopic,malevanets2000solute,malevanets2004mesoscopic}. The two parts of the algorithm are the following:\\
- the nonreactive part of the algorithm performs stochastic rotations of all fluid particle velocities to mimic the effect of collisions. To this end, a stochastic rotation operator $\omega_{\xi}$ is applied to all fluid particles located within the same cubic sub-region of the fluid. Specifically, the post-collision velocity ${\mathbf v^{'}}_{\xi,i}$ of a particle $i$ within the cubic sub-region of edge $a$ (denoted by $\xi$), is related to its pre-collision velocity $v_{\xi,i}$ via
\begin{equation}
{\mathbf v}^{'}_{\xi,i}= {\mathbf V}_{\xi} + \omega_{\xi} \left[ {\mathbf v}_{\xi,i} - {\mathbf V}_{\xi}\right]
\end{equation}
in which $\mathbf{V}_{\xi}=\frac{1}{N_\xi} \sum_{i=1}^{N_{\xi}} \mathbf{v}_{\xi,i}$ is the center-of-mass velocity for the $\xi$ region and the operator $\omega_{\xi}$ corresponds to a clockwise rotation of an angle $\alpha$ around an unit vector $\mathbf{n}_{\xi}$ with a random orientation,\\
- the reactive part of the algorithm is applied after the nonreactive part and consists of stochastic identity changes for the fluid particles. These identity changes mimic the outcome for the chemical reactions taking place in the fluid and which account for the self-propulsion of the nanomotor.\\
The algorithm so obtained allows for the conservation of momentum and kinetic energy. In addition, transport coefficients can be readily obtained through explicit Green-Kubo relations~\cite{kapral2008multiparticle}.\\

Alternative hydrodynamic simulation approaches include the lattice Boltzmann (LB) method~\cite{mcnamara1988use,dunweg2009lattice} and the dissipative particle dynamics (DPD) methods~\cite{espanol1995statistical}. The former couples a system of particles, standing for the self-propelled particle, to a lattice-Boltzmann model representing the solvent. Self-propelled particles, also termed as squirmers~\cite{kuron2019lattice,alarcon2013spontaneous,lintuvuori2016hydrodynamic,de2016lattice,bahr2021lattice}, are coarse-grained into an arrangement of mass points with a frictional coupling to the solvent and appropriate boundary conditions are applied at surfaces.
Instead of solving directly the Stokes equations (Eq.~\ref{STK}), the LB method solves the Boltzmann transport equation which obeys the same conservation laws and describes the evolution of the single-particle phase space probability distribution $f({\mathbf r}, {\mathbf v}, t)$, {\it i.e.} the probability of finding a fluid molecule with velocity ${\mathbf v}$ and position ${\mathbf r}$ at time $t$. As discussed by Kuron {\it et al.}~\cite{kuron2019lattice}, the LB method linearizes the relaxation of $f$ to the Maxwellian equilibrium. This is achieved by discretizing space on a cubic lattice and by discretizing time. The probability is allowed to flow between neighboring cells through a finite set of velocities ${\mathbf c}_i$. Fluid-particle interactions take place exclusively via boundary conditions, with no-slip boundary conditions introduced through reflections of the populations streaming into the boundary back into the fluid. Momentum transfers between particles and fluid are then accounted for by these reflections~\cite{kuron2019lattice}, leading to the following force ${\mathbf F}_{bb}(t)$
\begin{equation}
{\mathbf F}_{bb}(t)=a^3\sum_{{\mathbf r}_b} \sum_i {\mathbf c}_i \left( f_i ({\mathbf r}_b,t) - f_i ({\mathbf r}_b - {\mathbf c}_i \tau,t)    \right)
\end{equation}
and torque ${\mathbf T}_{bb}(t)$
\begin{equation}
{\mathbf T}_{bb}(t)=a^3\sum_{{\mathbf r}_b} \sum_i ({\mathbf r}_b - {\mathbf r}) \times {\mathbf c}_i \left( f_i ({\mathbf r}_b,t) - f_i ({\mathbf r}_b - {\mathbf c}_i \tau,t)    \right)
\end{equation}
in which ${\mathbf r}_b$ denotes a boundary (particle) node, $a$ the grid spacing for the lattice discretization and $\tau$ the time step for the discretization.

DPD is a multi-particle method that is akin to molecular dynamics, but with pairwise momentum-conserving stochastic and friction forces, and has been recently applied to study the collective properties of self-propelled particles~\cite{tao2009self,lugli2012shape,wang2019defect,wang2021enhancing}. In the DPD approach~\cite{espanol1995statistical}, the dynamics of the particles is given by the Langevin equations through the following stochastic differential equations
\begin{equation}
\begin{array}{ccc}
d{\mathbf r_i} & = & {{\mathbf p}_i \over m_i} dt\\
\end{array}
\end{equation}
\begin{equation}
\begin{array}{ccc}
d{\mathbf p_i} & = & \left[ { \sum_{j\ne i}{\mathbf F}^C_{ij}({\mathbf r}_{ij}) + \sum_{j\ne i} - \gamma \omega_D (r_{ij}) ({\mathbf e}_{ij} \cdot {\mathbf v}_{ij}){\mathbf e}_{ij} } \right] dt + \sigma \omega_R (r_{ij}) dW_{ij} \\
\end{array}
\end{equation}
where ${\mathbf r}_{ij}={\mathbf r}_i - {\mathbf r}_j$, $r_{ij}=|{\mathbf r}_{ij}|$ and ${\mathbf e}_{ij}= {\mathbf r}_{ij}/r_{ij}$ is the unit vector from particle $j$ to particle $i$. In the second equation, ${\mathbf F}_c$ denotes the conservative interparticle forces, while the second term on the right-hand-side corresponds to the dissipative forces and the third term to a Gaussian white-noise term, with $dW_{ij}$ as independent increments of a Wiener process. The functions $\omega_D$ and $\omega_R$ are weight functions that quantify the range of interaction for the dissipative and random forces, and $\gamma$ and $\sigma$ are the friction coefficient and amplitude of the noise.

Finally, Langevin dynamics in the overdamped limit has also been used to model nanomotors~\cite{gaspard2018fluctuating,gaspard2019thermodynamics}. In this case, a nanomotor $i$ is characterized by its position ${\mathbf r}_i (t)$ and axis $\hat{\mathbf {\nu}_i}(t)=(\cos \theta_i (t), \sin \theta_i (t))$ and obeys the following equations of motion
\begin{equation}
\begin{array}{lll}
\partial_t{\mathbf r}_i & = & v_0 \hat{\mathbf {\nu}_i} + \mu {\mathbf F}_i + {\mathbf \eta}^T_i (t)\\
\partial_t \theta_i & = & {\mathbf \eta}^R_i (t)\\
\end{array}
\end{equation}
Here, $v_0$ denotes the self-propulsion velocity, $\mu$ the mobility, ${\mathbf F}_i$ the force exerted on $i$, ${\mathbf \eta}_i^T (t)$ and $\eta^R_i (t)$ are translational and rotational white noise terms, respectively. This model was, for instance, recently applied to simulate the autonomous detection and repair microscopic mechanical defects and cracks by self-propelled Au/Pt nanomotors~\cite{li2015self}.\\

{\bf Simulation-aided design of nanomotors}\\

Responsive and active soft materials have drawn considerable attention over the past decade~\cite{balazs2010emerging,liu2020responsive,walther2020responsive}. In particular, synthetic and biomimetic nanomotors could potentially revolutionize nanomedicine and nanorobotics. Theories and simulations now play an active role in the improvement of design and control strategies of nanomachines. For instance, coarse-grained simulations have been instrumental for the optimization of DNA nanotechnology. Ouldridge {\it et al.} proposed a new model for a two-footed DNA walker, designed to step along a reusable track. Applying a moderate tension to the track can provide a bias for the walker to step forward, but also help him recover from undesirable overstepped states. Moreover, these authors showed that the process by which spent fuel detaches from the walker strongly influences the motion of the walker along the track, and suggested several modifications to the walker to improve its operation~\cite{ouldridge2013optimizing}. Chen {\it et al.} found a novel way to characterize the swimming motion of a linear catalytic nanomotor in a 2D fluid~\cite{chen2011characterizing}. The diffusion of the nanomotor was accelerated by the chemical propulsion as long as they confined its rotational degree of freedom. They also suggested how, in experiments, analyzing the confined diffusive behavior of nanomotors collectively could prove more efficient than tracking the trajectory of each nanomotor individually. Simulations can also help enhance the capacity of experimental synthetic nanomachines. Ortiz {\it et al.}~\cite{ortiz2016convective} studied how simulations could help design and harness enzymatic catalysis for on-demand pumping in nanofluidic devices. In particular, they modeled urease-based pumps and identified novel spatiotemporal variations in pumping behavior, thereby suggesting how self-powered fluidic devices, based on enzymatic pumps, could be improved.\\

Simulations can also serve as {\it in silico} experiments to advance applications in nanomedicine that rely on self-healing nanomaterials and targeted drug delivery. Li {\it et al.} ~\cite{li2015self} developed an nanomotor-based autonomous repair system that sought and localized cracks and mimicked wound healing. Nanomotors were obersved to form``patches'' and repair scratched electrodes, thereby restoring the conductive pathway. Fluid pumps have also been an area of intense research. For instance, Tansi {\it et al.} presented a new method to construct, move, and organize particle islands using light-powered fluid pumping~\cite{tansi2019organization}. Their method relied on freely suspended nanoparticles to generate fluid pumping towards desired point sources. The pumping rates were found to depend on particle concentration and light intensity, making them easy to control. Molecular dynamics simulation can also help design nanomachines as drug delivery systems. Recently, Cai {\it et al.} carried out molecular dynamics simulations to help design a new rotary nanomotor for a drug delivery nanosystem, involving graphene origami to drive a carbon nanotube rotor~\cite{cai2021carbon}. Such screening methods can be of great significance when testing the different components of a nanomachine.\\

Physical models were proposed recently to improve our understanding of the self-assembly and collective motion of nanoobjects. In particular, approaches taking into account both the hydrodynamics and the chemistry are crucial to elucidate the behavior of nanomotors propelled by self-diffusiophoresis~\cite{robertson2018synthetic}. The inhomogeneous concentration fields induced by asymmetric motor reactions are ``felt'' by other motors and, as a result, strongly influence their motion. Systems composed of a collection of Janus particles can, for instance, exhibit dynamic cluster states~\cite{liebchen2015clustering}. Particles can join, leave or be trapped within a cluster~\cite{ginot2018aggregation,liebchen2017phoretic}. Prior work has shown the importance of the catalytic capsize, motor density, interaction potentials and fluid properties in cluster formation~\cite{huang2017chemotactic}. Models also provide insight into forward- and backward-moving motors. For sphere-dimer motors, forward-moving motors are attracted towards the areas with high product concentrations and, as a result, move toward other motors, while backward-moving motors tend to avoid other motors. The flow field is also found to depend on the bond length in the dimer, with forward-moving motors acting as pullers for short bond lengths and as pushers for long bond lengths~\cite{thakur2012collective, colberg2017many}. Chemical oscillations can also impact the collective behavior. Motors show little tendency to cluster where fuel concentration is low, while they form dynamic clusters where fuel concentration is high~\cite{robertson2015nanomotor}. Other studies have assessed the ability of chemically propelled nanomotors to perform tasks in complex media crowded by obstacles of various kinds. For instance, in nature, molecular machines can carry out diverse biochemical and transport tasks by moving on biofilaments or operating in membranes. Simulations by Qiao {\it et al.}~\cite{qiao2020active} on oligomeric motors attached to a filament provide insight into the self-generated concentration fields produced by the catalytic reactions.\\ 

{\bf Controlling the motion of nanomotors}\\

Brownian motion is a crucial feature at the nanoscale and considerably increases the difficult to control nanosized objects. With the advent of molecular nanotechnology, increasing effort has focused on the production of the tiniest lego possible~\cite{peplow2015tiniest}. Improvements in design and synthesis of molecular machines, such as nanoscale motors, rotors, switches, and pumps, have led to tremendous advances~\cite{stoddart2017mechanically,kistemaker2021exploring,sauvage2017chemical}. However, Brownian motion presents a real problem for the design and the manufacture of molecular-scale machines and factories~\cite{kay2007synthetic}. Recently, there has been a paradigm shift, according to which Brownian motion can be seen as an unexpected help rather than a disadvantage. For instance, Toyabe {\it et al.}~\cite{toyabe2010experimental} designed smart devices that could power themselves using Brownian motion. More specifically, they showed that dimeric particles could rotate clockwise by converting information into energy through a feedback manipulation of Brownian motion. Millen {\it et al.} proposed a method utilizing Brownian motion to measure the temperature of nanoscale objects~\cite{millen2014nanoscale}. In this case, Brownian motion results from the collisions with the surrounding gas (O$_2$ and N$_2$) molecules. The surface temperature of the nanoobject can then be inferred from the collision features. This new procedure is thus an exceptional opportunity to better operate and control nanoscale systems. Indeed, it opens the door to the use of thermal energy as a lever for fine-tuning their activity. Microscopically, the temperature is often calculated from the kinetic energy stored in the velocity degrees of freedom of atoms or molecules. In the case of colloidal particles suspended in a fluid (overdamped motion), the kinetic energy is constantly dissipated into heat. This heat is then turned back into motion via the fluctuations of the hydrodynamic velocity field of the solvent. Fluctuation and dissipation relations, such as, {\it e.g.}, the Stokes-Einstein relation, give a route for the calculation of the temperature and transport properties of the fluid from the Brownian fluctuations of suspended probes. “Hot Brownian motion” recently emerged as a new non-isothermal out-of-equilibrium concept. This is the case, for instance, when various degrees of freedom of the particle ({\it i.e.}, for a sphere, translational and rotational positions, and momenta) are each predicted to have their own effective temperatures. This departure from the equipartition principle is evidence that the system is very far from equilibrium~\cite{rings2012rotational,falasco2014effective, kroy2016hot}. Schachoff {\it et al.} used heated gold nanoparticles to reveal the impact of a radially symmetric temperature profile around the nanoparticle. Their results show that an effective temperature and the friction properties of a hot Brownian motion can be defined in terms of a fluctuation-dissipation relation similar to that of isothermal systems. Moreover, asymmetric temperature profiles were found to induce a self-thermophoretic propulsion and how the use of DNA origami coupled to the nanoparticle could allow for the control of rotational motion~\cite{schachoff2015hot} Environmental feature can also outweigh the increased rotational diffusion of nanoswimmers. Wu {\it et al.} reported an anomalously rapid transport of self-propelled nanoswimmers in a porous matrix~\cite{wu2021mechanisms}. In addition, they showed that nanoswimmers escaped from cavities more than an order of magnitude faster than expected, when compared to the corresponding Brownian particles. Moreover, self-propulsion resulted in qualitatively different phenomena, such as surface-mediated searching and the cancellation of energy barriers at hole exits~\cite{stark2016swimming}. Finally, recent reports show that collective effects and emergence can lead to a controllable directed motion~\cite{saha2014clusters}.

\section{Active fluids at the microscale}

\subsection{Real-life Systems}

{\bf Microbiology}\\

Microbiology studies biological microorganisms, such as bacteria, viruses, and protozoa. Famous microbiologists include, but are not limited to, Jenner and his vaccine against smallpox and Fleming with the discovery of penicillin. Microbiology is a fascinating field that explores life diversity on Earth and the existence of life elsewhere in the Universe. Many microorganisms have yet to be discovered. It is estimated that there are one trillion microbial species on Earth and that 99.999 percent of microbial species have not been identified~\cite{locey2016scaling,shoemaker2017macroecological}. Many microorganisms fulfill important tasks by helping make drugs, manufacturing biofuels, and cleaning up pollution. Unicellular swimmers such as, {\it e.g.}, the {\it Escherichia coli} bacteria, are typically of a few to several tens of micrometers in size. Because these microswimmers live in a world where the viscous forces are greater than the inertia forces (low Reynolds number), microorganisms have refined over time their propulsion strategies, which successfully overcome and even utilize viscous drag~\cite{berg1975bacteria}. Moreover, microswimmers hardly ever swim alone. Indeed, in assemblies of motile microorganisms, cooperativity reaches a new level of complexity as they exhibit highly organized movements with remarkable large-scale patterns such as networks, complex vortices, or swarms.~\cite{plesset1974bioconvection}\\

{\bf Bacteria and algae}\\

\begin{figure}[ht]
\centering
\includegraphics[width=9cm]{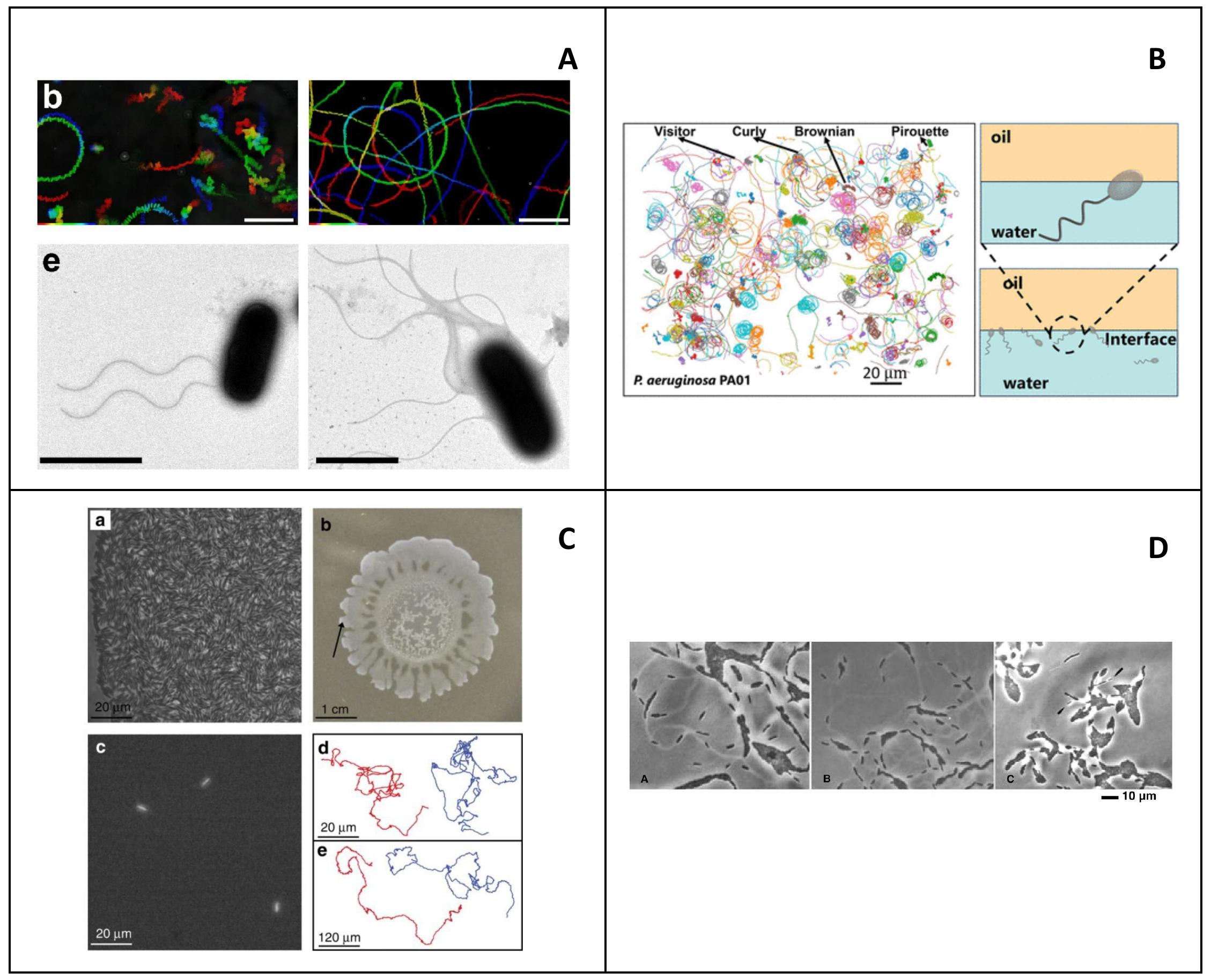}
\caption{A) (Top) Swimming motility and structural parameters of {\it E. coli} ATCC10798 and W3110 through sequential phase-contrast images~\cite{kinosita2020distinct}. Images are taken at $50$~ms intervals for $10$~s and integrated with an intermittent color code: “red $\rightarrow$ yellow $\rightarrow$ green $\rightarrow$ cyan $\rightarrow$ blue.” Scale bar, 20 $\mu$m. (Bottom) Electron micrographs of {\it E. coli} ATCC10798 (left) and W3110 (right) cells. Scale bars, 2 $\mu$m. Adapted with permission from {\it Sci Rep.}~ {\bf 2020}, 10, 15887. Copyright 2020 Springer Nature. B) Ensemble behavior of {\it Pseudomonas aeruginosa PA01} at oil-aqueous interfaces~\cite{deng2020motile}. Individual bacteria trajectories are shown only over a 20~s time span for clarity, with active motions of (i) interfacial visitors, persistent circular motions including (ii) pirouettes and (iii) curly trajectories. A fourth trajectory type (iv), inert Brownian diffusive bacteria, is also present. Reprinted with permission from {\it Langmuir}~ {\bf 2020}, 36, 6888-6902. Copyright 2020 American Chemical Society. C) Swarming bacteria migrate via a Levy Walk, with in (a-b), a phase contrast imaging of a {\it B. subtilis} swarming colony, in (c), fluorescent microscopy showing the fluorescently labelled bacteria, and in (d-e) example trajectories of individual bacteria inside the swarm at high (d) and low (e) magnifications~\cite{ariel2015swarming}. Reprinted with permission from {\it Nat. Commun.}~ {\bf 2020}, 6, 8396. Copyright 2015 Nature Springer. D) Slime trails~\cite{wolgemuth2002myxobacteria}. Deposition of slime by {\it M. xanthus} as it glides on agar. (A) A$^+$S$^+$ strain DK1622. (B) A$^+$S$^-$ strain DK10410. (C) A$^-$S$^+$ strain ASX1. Photographs
of the swarming edge were taken after 1 day. Reprinted with permission from {\it Curr. Biol.}~ {\bf 2002}, 12, 369-377. Copyright 2002 Elsevier.}
\label{fig4}
\end{figure}

Bacteria use a system of helical filaments called flagella for propulsion. Different bacteria have different arrangements of flagella depending on what they need to achieve in terms of motility. They can have either a single flagellum or multiple flagella located at a particular spot on their surface. Alternatively, there can be a single flagellum on each of the two opposite ends or multiple flagella pointing in all directions~\cite{janssen2011coexistence}. An example of the latter is {\it E. coli}, shown in Fig.~\ref{fig4}A. In this case, the flagella arrangement allows for a very interesting swimming pattern, known as "run-and-tumble" motion~\cite{macnab1977normal,macnab1977bacterial,berg2004coli}. During the "run" phase, flagella form a bundle (counterclockwise rotation), that pushes the bacterium forward in one direction. In the "tumble" phase, one or a few flagella leave the bundle (clockwise rotation), which leads to a random change in the orientation of the bacterium. The frequency of each of these two steps informs on the bacterium's local environment. For instance, the frequency of the ``run phase'' correlates with how favorable the local environment. If there are nutrients close to the bacterium, the bacterium will undergo a rapid forward motion to access the nutrients. When there are no nutrients, the bacterium undergoes the ``tumble'' phase and starts looking for nutrients elsewhere. The run-and-tumble motion is thus akin to a goal-oriented navigation. In this case, the bacterium reacts to a chemical gradient in nutrients through chemotaxis. The bacteria may also react to other stimuli including temperature changes (thermotaxis), pressure changes (barotaxis), and flow changes (rheotaxis). In addition to bacteria, algae such as volvocine green algae have emerged as model organisms for flagellar propulsion. In particular, Volvox has allowed for an improved understanding of the transition from unicellular to multicellular life. This multicellular green alga forms spherical colonies of up to 50,000 cells. The cells move their flagella in a coordinated fashion, which enables the colony to swim, for example, towards light~\cite{solari2011flagellar,herron2016origins}.\\

{\bf Swarming and gliding}\\

Bacteria can also swim as a group and exhibit collective motion (see Figs.~\ref{fig4}B~and~C). For instance, the formation of swarms of bacteria has been reported close to a moist surface or in thin liquid films~\cite{copeland2009bacterial,zhang2010collective,ariel2015swarming,harshey1994bees}. Contrary to swimming, swarming implies specific changes in cell shape, with the cells becoming more elongated through the suppression of cell division. Swarms also lead to the formation of a new entity with a very large number of flagella. This points to a significant role of flagella and flagella interactions between adjacent cells in the swarming process~\cite{stahl1983extracellular,shapiro1998thinking,jones2004ultrastructure,kearns2010field,verstraeten2008living,lauga2016bacterial}. Myxobacteria (see Fig.~\ref{fig4}D) are other examples of bacteria traveling in swarms. They exhibit a different form of bacteria motility, known as gliding. In this case, cells move on a substrate (or through a gel or porous material). For instance, {\it Myxococcus xanthus} move many cell lengths over surfaces without flagella, giving rise to rippling patterns. Genetic and cell behavioral studies have identified two motion patterns for the gliding motion of {\it M. xanthus}. These two patterns are governed by different genes, corresponding to A (for adventurous) and S (for social) cell behaviors. In the former, $A^{+}$ stands for 'A-motility' for which single cells move, resulting in a spatial distribution with many single cells. $S^{+}$ stands for S-motility, in which isolated cells do not move, but cells close to one another undergo motion~\cite{kaiser1979social}. While $A^{-}S^{-}$ strains are nonmotile and never move more than $1/4$ a cell length, both $A^{+}S^{-}$ and $A^{-}S^{+}$ strains are motile. However, their swarm patterns and swarming rates differ from $A^{+}S^{+}$~\cite{nan2011uncovering,wolgemuth2002myxobacteria,mauriello2010gliding,munoz2016myxobacteria}. This demonstrates how bacteria adapt and leverage two propulsion systems in a synergistic way.

\subsection{Synthetic Systems}

{\bf Active colloids and Janus micromotors}\\

Microtechnology has undergone tremendous progress in recent years. It is now possible to design and synthesize microstructured materials with dimensions matching the size of a cell or collections of cells. These structures are exceptional tools as they offer the possibility to control the interface between cells and their interactions with their chemical and physical environment. Exciting developments in this field stem from soft lithography named for its use of soft, elastomeric elements in pattern formation~\cite{xia1998soft,rogers2005recent}. Soft lithography enables printing, molding, and embossing using an elastomeric stamp with feature sizes ranging from 30 $nm$ to 100 $\mu m$. Recent advances include the design of three-dimensional curved structures, the ability to work with different materials, and the generation of a well-defined and controllable surface chemistry. Soft lithography can yield channel structures appropriate for microfluidics, as well as pattern and manipulate cells~\cite{whitesides2001soft,qin2010soft}.

Several sophisticated micromotors were synthesized in recent years. Helical microswimmers were developed for targeted therapies~\cite{wang2017bioinspired}, environmental sensing~\cite{gao2014bioinspired} and monitoring, cell manipulation and analysis, and lab-on-a-chip devices. Qiu {\it et al.} proposed the first functionalized artificial bacterial flagella. These 3D microswimmers can deliver plasmid DNA into targeted cells using rotating magnetic fields. Cells targeted by f-ABFs were successfully transfected by the transported pDNA and expressed the encoding protein~\cite{qiu2015magnetic}.Patchy particles are an emerging tool for the synthesis of intelligent, structured micro-objects. Such colloidal particles can be anisotropically patterned, either by surface modification (localized attractive spots) or through their shape. Janus particles are unique among these micro-objects as they can have different chemical or physical properties and directionality within a single particle. In particular, the 'two-faced' spherical Janus particles play a significant role as micromotors~\cite{walther2013janus,yi2016janus}. Active Janus colloids are capable of propelling themselves in fluidic environments via localized and asymmetric catalytic reactions that decompose, for instance, hydrogen peroxide~\cite{sanchez2015chemically,wang2015fabrication}. Gao {\it et al.} proposed catalytic iridium-based Janus micromotors that only require 0.001\% of the chemical fuel to self-propel at 20 body lengths s$^{-1}$. In this case, Janus micromotors are composed of Ir and SiO$_2$ hemispheric layers. The catalytic decomposition of hydrazine at the Ir interface creates propulsion through osmotic effects. Such a low fuel concentration represents a 10,000-fold decrease in the level required for catalytic nanomotors~\cite{gao2014catalytic}. Janus micro-objects can also serve as a building block for selective functionalization in biomedical applications. Wu {\it et al.} developed an autonomous self-propelled Janus capsule motor that can also serve as smart cargo~\cite{wu2012autonomous}. This capsule motor is composed of partially coated dendritic platinum nanoparticles on one side, allowing for the catalytic decomposition of hydrogen peroxide and the generation of oxygen bubbles. This process then recoils the motion of the capsule motors. The capsules can autonomously move at 125 body lengths/s while exerting large forces exceeding 75 pN. Finally, these asymmetric hollow capsules can achieve directed motion using an external magnetic field. Recently, Janus micromotors have been applied to water treatment~\cite{jurado2015self,dong2017visible}, and analytical sensing~\cite{jurado2017magnetocatalytic,sentic2014electrochemiluminescent}.\\ 
{\bf Modular microswimmers and directed self-assembly}\\

Can we construct artificial microswimmers from different components? Differently put, is it possible to create modules combining different functions and assemble them altogether? Recent research has started to address this challenge~\cite{niu2018modular}. For instance, autonomous microswimmers, which include active and inactive functional components, have been assembled to create self-propelling complexes. The resulting modular microswimmers can exhibit different types of modular swimming. More generally, two kinds of modular swimmers were designed in recent years. The first is called swimmers with bound structures. In this case, chemical bonds or electrostatic interactions link all components, thus limiting the rotational degree of freedom. Dreyfus {\it et al.} created a flexible artificial flagellum with a linear chain of colloidal magnetic particles linked by DNA and attached to a red blood cell, and uses an external uniform magnetic field to align the filaments. Oscillating a transverse field then allowed for the actuation of the movement, thereby inducing a beating pattern that propelled the structure~\cite{dreyfus2005microscopic}. Colloidal rotors ~\cite{tierno2008controlled}, self-propelled sphere dimers~\cite{valadares2010catalytic}, a colloidal chain made of Janus particles with a zigzag-shaped arrangement to form rotators~\cite{vutukuri2017rational}  and magnetic microlassos for cargo delivery~\cite{yang2017magnetic} were also recently designed. The second type of modular swimmer is called dynamic structures. In this case, the composition and organization of components can change dynamically, as the modular swimmers can rearrange, disassemble and re-assemble in response to external fields or self-generated gradients. For instance, Snezho {\it et al.} designed a self-propelling snake from a dispersion of magnetic microparticles at a liquid-air interface that is energized by an alternating magnetic field~\cite{snezhko2009self}. Helical ribbons have been self-assembled as paramagnetic beads in an external magnetic field~\cite{casic2013propulsion}. An external periodic magnetic field also served to create throwers and rowers from asymmetric paramagnetic beads~\cite{vilfan2018magnetically}. Other fascinating examples are reconfigurable microswimmers. Du {\it et al.} designed a two-body swimmer from by paramagnetic beads under an eccentric magnetic field~\cite{du2018reconfigurable}. Palacci {\it et al.} developed colloidal dockers from a peanut-shaped hematite, guided by a weak magnetic field to the vicinity of a colloid. The two then couple via a light-activated phoretic force induced by a chemical gradient~\cite{palacci2013photoactivated}. Electric fields can also be used to generate a rotating pinwheel formed by Janus particles around a homogeneous colloid under an AC electric field~\cite{zhang2016natural}. Asymmetric colloidal dimer can be propelled by an electrohydrodynamic flow~\cite{ma2015inducing}, and mobile microelectrodes with Janus particle can selectively attract or repel colloids by dielectrophoresis in a vertical electric field~\cite{boymelgreen2018active}. An alternative strategy, based on self-generated fields, can be used for the self-assembly of dynamic modular swimmers. This is the case for autonomous movers capable of gliding across the surface of a liquid without an external power source~\cite{ismagilov2002autonomous}. For such systems, the motion of individual objects was powered by the catalytic decomposition of hydrogen peroxide, while self-assembly resulted from capillary interactions at the fluid/air interface~\cite{ismagilov2002autonomous}. Different types of interactions can play a significant role in the self-organization of micromachines. For instance, two tadpoles can self-assemble into a cluster bound by van der Waals interaction~\cite{gibbs2010self}. Janus micromotors and a non-catalytic hydrophobic colloid can couple via hydrophobic interaction~\cite{gao2013organized}. Catalytic reactions can also give rise to hydrodynamic forces that self-assemble microrotors and swimmers~\cite{wykes2016dynamic}. Martinez {\it et al.} use reconfigurable photoactivated magnetic microdockers to assemble and transport microscopic cargos~\cite{martinez2017assembly}. In the latter, the photoactivation process induces a phoretic flow capable of attracting cargos toward the surface of the propellers. At the same time, a rotating magnetic field is used to transport the composite particles to any targeted location. The method allows for the assembling of small colloidal clusters of various sizes, composed of a skeleton of mobile magnetic dockers, which cooperatively keep, transport, and release the microscopic cargos. Modular phoretic microswimmers were also designed using a concentration gradient induced by ion exchange.~\cite{niu2017assembly} and from heat-induced phoretic forces~\cite{schmidt2019light}.\\

\begin{figure}[ht]
\centering
\includegraphics[width=10cm]{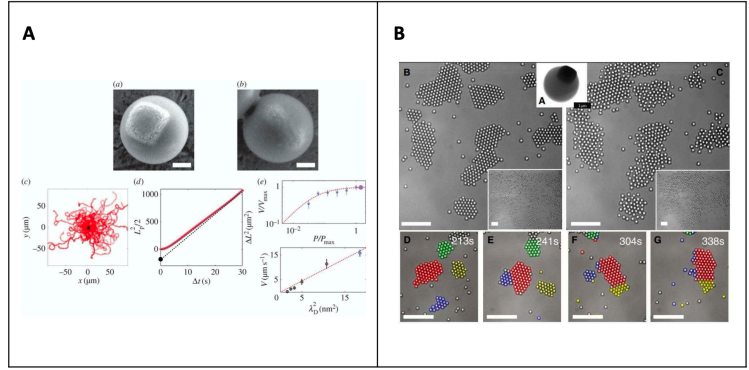}
\caption{A) Light-activated self-propelled colloids with (a-b) Scanning electron microscope (SEM) images of hematite particles embedded in a TPM (3-(Trimethoxysilyl)propyl methacrylate) shell, (c) Trajectories of colloidal microswimmers in the absence of light (black) or with the light on (red), and (d) Averaged mean square displacement (red symbols), compared to a random-walk dynamics (blue dashed line), (e) self-propulsion velocity against light intensity for blue light (blue symbols) and UVA-violet light (violet symbols), and with a red dashed line corresponding to a fit with a Michaelis–Menten kinetics, and (f) Self-propulsion velocity against the Debye length of the solution, varied by addition of sodium chloride salt (black symbols) and withdrawing of the surfactant (blue symbol)~\cite{palacci2014light}. Reprinted with permission from {\it Phil.Trans. R. Soc. A}~ {\bf 2014}, 372, 20130372. Copyright 2014 The Royal Society. B) Living crystals of light-Activated colloidal swimmers, with in (B.A) TPM polymer colloidal sphere and protruding hematite cube (dark). Living crystals (B.B) are assembled from a homogeneous distribution (inset) when a blue light is switched on, and melt (B.C) by thermal diffusion when the light is off. (B.C) shows the system 10s after light is turned off (inset, after 100s). (B.D to B.G) Colors indicate the time evolution of particles belonging to different clusters. In (B.B-B.G), scale bars correspond to 10$mm$~\cite{palacci2013living}. Reprinted with permission from {\it Science}~ {\bf 2013}, 339, 936-940. Copyright 2013 American Association for the Advancement of Science.}
\label{fig5}
\end{figure}

{\bf Microswimmers in complex environments}\\

While recent studies have shed light on propulsion mechanisms at the microscale, a complete understanding of the dynamics of microswimmers in complex environments still eludes us~\cite{bechinger2016active}. For instance, micromotors travel in mucus gels, blood vessels, and microfluidic chips to perform their tasks. These complex environments can be seen as various types of confinements~\cite{xiao2018review}. The transport dynamics of micromotors depends on the geometry of the confinement. In addition to thermal fluctuations and self-propulsion, the behavior of confined microswimmers is also impacted by interfaces. Studies have indicated a strong coupling between the microswimmers' motion and interfaces, leading to the intriguing transport behaviors near interfaces~\cite{wu2021mechanisms}. For example, in their study of spherical Janus colloids close to solid surfaces, Das {\it et al.} reported an active quenching of the particles' Brownian rotation, which leads to constrained in-plane swimming along the wall. This new steering mechanism leads to a constrained 2D-enhanced diffusion at the walls~\cite{das2015boundaries}. Here, the main contribution comes from the dynamic flow field at the interface, but not from the interactions between the wall and the particles. Other studies have shown how directed motion can be induced by combining geometric constraints. Another example is the study by Brown {\it et al.} of catalytic Janus particles swimming and hopping between colloids in a two-dimensional colloidal crystal. The hopping rate was found to vary inversely with fuel (hydrogen peroxide) concentration~\cite{brown2016swimming}. This has led to new approaches to achieve directional guidance of chemically active microswimmers. Simmchen {\it et al.} used various topographic features (stripes, squares, or circular posts) to guide the motion of Janus microswimmers~\cite{simmchen2016topographical}. Microswimmers followed step-like topographical features that were only a fraction of the particle radius in height.

Optical traps can also be used to confine microswimmers~\cite{liu2021opto,jahanshahi2020realization}. Nedev {\it et al.} deisgned the first microswimmer "elevator"~\cite{nedev2015optically} by trapping Janus particles, with a silica sphere with a gold half-shell, with optical tweezers and by moving them along the axis of the laser beam, thereby allowing for the upward and downward motions of the Janus particles. They showed that this process arose from a complex interplay between optical and thermal forces, with scattering forces orienting the asymmetric particle, while the strong absorption on the metal side inducing a thermal gradient, resulting in particle motion. Thus, an increase in laser power led to upward motion, while a decrease in laser power resulted in downward motion. Gao {\it et al.} performed an angular trapping of Janus particles by controlling the laser polarization direction~\cite{gao2020angular}. Optical tweezers were also used to trap self-propelled Janus particles in a "round-trip" motion~\cite{liu2016self} and trochoidal trajectories~\cite{moyses2016trochoidal}. However, hematite-based microswimmers are generally ejected from conventional optical tweezers. Recently, Abacousnac {\it et al.} develop dark optical tweezers as holographic optical trapping systems. They were able to trap dielectric spheres enclosing a hematite cube (dark-seeking particles) and move them alomg the three dimensions~\cite{abacousnac2022dexterous}.\\

{\bf Toward living materials: collective motion in active fluids}\\

Active nematics constitute a new class of active fluids, which combine the properties of liquid crystals properties with those of active matter. Nematic liquid crystals are generally modeled as rod-like particles, mimicking the shape of elongated micro-objects. Depending on temperature or concentration, these elongated particles can predominantly align in a given direction, {\it i.e.}, in a nematic phase with long-range orientational. Structural inhomogeneities or the application of an external force can lead to mismatches between neighboring domains with different directions, resulting in topological defects and singularities in the orientation field~\cite{de1993physics}. There are two types of topological defects: comet-like $(+1/2)$ or trefoil-like $(-1/2)$ in 2D nematic liquid crystals, with the topological charge $+1/2$ or $-1/2$ standing for the angle by which the particles turn ($+180$ for a charge of $+1/2$ and $-180$ for a charge of $-1/2$). Cortes {\it et al.} have shown that topological defects in active nematic liquid crystals can be annihilated in pairs $(+1/2$ and $-1/2$) or created in pairs ($+1/2$ and $+1/2$, or $-1/2$ and $-1/2$)~\cite{cortese2018pair}. Bacteria can be mimicked by adding activity to elongated microparticles, making them self-propelled. This autonomous motion can be viewed as a force capable of creating/annihilating topological defects. Several groups~~\cite{doostmohammadi2018active,giomi2014defect,decamp2015orientational} have established that topological defects could lead to the onset of complex streaming flows. Active turbulence is a chaotic-like feature that destroys the long-ranged nematic order. The challenge is thus to control the direction of the bacteria but also command their chaotic behavior. Zhou {\it et al.} used motile rod-shaped bacteria in water-based nontoxic liquid crystals to obtain living liquid crystals (LLCs)~\cite{zhou2014living}. They reported that the coupling between the long-range orientational order in the liquid crystal and the swimming activity of the bacteria dramatically altered both the individual and collective bacterial dynamics. For instance, the motion of the bacteria perturbs the orientational order of the liquid crystal, even resulting in local melting and allowing for a direct observation of the bacteria's motion. Collective motion in this active fluid also gives rise to self-organized textures unseen in equilibrium liquid crystals~\cite{zhou2014living}. Peng {\it et al.} have shown that active matter can be controlled via topological defects and patterns~\cite{peng2016command}. Orientational order in a liquid crystal can direct the flow of self-propelling bacteria which, in turn, impact the patterning of the liquid crystal molecules. Patterns on a substrate can lead to surface anchoring of the liquid crystals that results in the ordering of the bacteria. Bacteria were found to be able to differentiate between different types of topological defects, as bacteria headed toward defects of positive topological charge and avoided negative charges. Understanding of the interplay between hydrodynamics and topology will be key in the design, control, and manipulation of soft active matter for biosensing and biomedical applications.\\

Swarming and pattern formation are large-scale phenomena that demonstrate a form of intelligence as exhibited by bacteria when forming colonies. Palacci {\it et al.} studied the self-organized clustering of light-activated colloidal surfers~\cite{palacci2013living}. They demonstrated that they could control the formation of clusters of microsurfers by switching on and off blue light (see Fig.~\ref{fig5}). Palacci {\it et al.} were able to relate this observation to the general property known as giant-number fluctuations found in swarming systems~\cite{ginelli2016physics}. The microsurfers clusters were named two-dimensional "living crystals," as clusters formed, broke, exploded, and formed again elsewhere. More generally, in another type of dissipative system, Naryan {\it et al.} observed long-lived giant number fluctuations in a swarming granular nematic system, and showed that an agitated monolayer of rodlike particles could exhibit liquid crystalline order. This showed that the onset of flocking, coherent motion, and large-scale inhomogeneities could take place in a system in which particles did not communicate except by contact, had no sensing mechanisms, and were not impacted by the environment~\cite{narayan2007long}. Dissipative building blocks, such as photoactive components, can also be used to control self-assembly. Aubret {\it et al.} succeeded in the targeted formation and synchronization of self-powered microgears. These self-spinning microgears can follow spatiotemporal light patterns demonstrating the possibility to program interactions and direct self-assembly. It lays the groundwork for the autonomous construction of dynamical architectures and functional micro-machinery~\cite{aubret2018targeted}. Other reconfiguring self-assembly systems were designed by Kang {\it et al.}~\cite{kang2022reconfiguring}, who formed chains and flower structures by using photoresponsive hybrid colloids. They also succeeded in transforming the chains into flower-like structures with decreasing UV intensity and triggered the formation of orientationally ordered clusters by applying an external magnetic field. Several discoveries have recently paved the way for the control and reconfiguration of micro-objects self-assembly. This opens the door to the design of novel smart materials such as reconfigurable robots and programmable soft robotics swarms for cooperative grasping, collective cargo transportation, and the building of micro-factories~\cite{deng2018swarming,balazs2018intelligent,wang2015one}.\\

\subsection{Theory and Simulations}

{\bf Modeling across the scales: the Vicsek model}\\

The transition to collective motion is also often termed as flocking. In fact, it is commonly observed in a wide range of living systems~\cite{ginelli2016physics,schaller2010polar}. This includes small systems at the subcellular or cellular level~\cite{trepat2009physical}, bacterial colonies~\cite{zhang2010collective}, and large systems such as schools of fish, flocks of birds~\cite{parrish1997animal} and herds of mammals~\cite{ginelli2015intermittent}. A well-known model for active particles is the Vicsek model~\cite{vicsek1995novel}. This model captures the minimal ingredients for active particles to undergo a transition from single self-propelled particles to collective motion~\cite{vicsek1995novel,chate2008collective,chate2008modeling,solon2015phase,ginelli2016physics}. The Vicsek model accounts for the overdamped dynamics of a system composed of $N$ self-propelled particles that interact through a local alignment rule. Each particle $j$ is characterized by ${\mathbf r_j}$, which stands for its position, ${\mathbf v_j=v_0 {\mathbf{s_j}}}$ its velocity, $v_0$ the norm of the velocity, and ${\mathbf s_j}=(cos \theta_j, sin \theta_j)$ its heading, or direction of motion. Accounting for angular noise, the equations of motion for the active particles are given in 2D by 
\begin{equation}
\begin{array}{lll}
{\mathbf r_j}(t+\Delta t) & = & {\mathbf r_j}(t)  + \Delta t v_0 {\mathbf s_j}(t+\Delta t) \\
{\theta_j}(t+\Delta t) & = & Arg \left[ \sum_{k \in D_j}  e^{i \theta_k (t)} \right] +  N_{D_j} \eta \xi_j \\
{\mathbf v_j}(t+\Delta t) & = &v_0  e^{i{\theta_j}(t+\Delta t)}  \\
\end{array}
\label{VM-AN}
\end{equation} 
where the average heading at time $t$ is calculated over all neighboring particles $k$ within a unit disk $D_j$ around particle $j$, $\xi_j$ is an angle chosen randomly between $-\pi$ and $\pi$, $\eta$ denotes the noise intensity. The random angle $\xi_j$ is a zero-average, delta-correlated scalar noise termed as white noise because of its flat Fourier spectrum~\cite{ginelli2016physics}. Several studies have introduced variations on this model that include a short-range repulsion, or steric interactions, between particles~\cite{chate2008collective} or allow the norm of the self-propelling velocity $v_0$ to fluctuate~\cite{peshkov2014boltzmann}.\\ 

Several features characterize the Vicsek model and account for the transition to collective motion. First, particles are self-propelled. Second, they change their relative positions in a complex manner, {\it i.e.}, according to their velocity fluctuations, leading to a far-from-equilibrium behavior~\cite{ginelli2016physics}. Finally, there are no conservation laws apart from the conservation in the number of self-propelled particles, which means that there is no momentum conservation. Strictly speaking, the latter differs from what takes place in the case of microswimmers in a suspension. Indeed, momentum is transferred from the microswimmers to the surrounding fluids and hydrodynamic interactions are expected to play a significant role, especially for 3D suspensions~\cite{winkler2016low}. The Vicsek model exhibits a disordered gas-like phase (high external noise), microphase separation with propagating bands of high density (intermediate external noise) and a polar liquid (low external noise)~\cite{chate2008collective,chate2008modeling,ginelli2016physics}. Toner and Tu developed fluctuating hydrodynamic equations for the Vicsek model~\cite{toner1995long,toner1995long,czirok1997spontaneously,czirok2000collective,toner2005hydrodynamics,toner2012reanalysis} and showed via a dynamic renormalization group calculation that these polar flocks possess a long-range orientational order, even in 2D. Although this theory only applies to dilute, aligning, dry active matter~\cite{marchetti2013hydrodynamics,mahault2019quantitative}, its predictions include the concept of giant number fluctuations which is relevant to a wide range of active matter systems~\cite{ramaswamy2003active,chate2006simple,narayan2007long,zhang2010collective,ngo2014large,giavazzi2017giant,nishiguchi2017long,levis2019activity}. Indeed, for orientationally ordered phases of active matter, the variance of the number of particles in subsystems of increasing size increases faster than the mean. A detailed analysis of the phase transitions in the Vicsek model has highlighted its similarity with the liquid-gas transition~\cite{solon2015phase,trefz2017estimation}. This is in agreement with the solutions found using hydrodynamic equations for
flocking models~\cite{caussin2014emergent,ihle2011kinetic,ihle2013invasion}.

Partial differential equations for the density field $\rho(x,t)$ and the momentum field $W(x,t)=\rho(x,t)P(x,t)$ in which $P(x,t)$ is a polarization field (between $0$ and $1$), can be written as 
\begin{equation}
\begin{array}{lll}
\partial_t \rho & = & - \partial_x W \\
\partial_t W & = & - \xi W \partial_x W + a_2 W - a_4 W^3 - \lambda \partial_x \rho + D \partial_{xx} W\\
\end{array}
\end{equation}
These equations account for a continuous mean-field transition from a homogeneous isotropic state with $\rho=\rho_0$ and $P=0$ when $a_2<0$ to a homogeneous polarized state with $P=\rho_0^{-1}\sqrt(a_2/a_4)$ when $a_2>0$.
The $\lambda$ term corresponds to the pressure gradient induced by density heterogeneities, while $\xi$ and $D$ are transport coefficients for the advection and the diffusion of the local order parameter~\cite{caussin2014emergent}. After integration~\cite{solon2015phase}, the solutions exhibit traveling bands with both smectic microphases and phase-separated profiles. We finally add that extensions of the Vicsek model to binary mixtures have involved systems with two sub-populations with different external noises~\cite{ariel2015order}, and mixtures of passive and active particles~\cite{martinez2018collective}.\\

{\bf MIPS at the micron scale}\\

Following the success of the Vicsek model, the next step has consisted in developing a minimal model for the simulation of wet active matter. This has led to the proposal of the Active Brownian Particles (ABP) model~\cite{romanczuk2012active,stenhammar2013continuum,stenhammar2014phase,solon2015active,ebeling1999active,cates2013active,speck2014effective,ganguly2013stochastic,turci2021phase,volpe2014simulation,marconi2017heat,caporusso2020motility,schwarz2012phase,hagan2016emergent,mallory2018active}. This model quickly proved to be instrumental in furthering our understanding of active matter. Indeed, it provided a testing ground for the hypothesis that clustering and phase separation were intrinsic properties of active systems~\cite{tailleur2008statistical,cates2010arrested}, and resulted from the flux of chemical energy that drives motility and breaks detailed balance~\cite{cates2012diffusive}. Fily and Marchetti~\cite{fily2012athermal,marchetti2016minimal,fily2014freezing,henkes2011active} used an ABP model to show that clustering could be observed even in the absence of any alignment rule, unlike in the Vicsek model. In the ABP model, particles are modeled as soft repulsive disks, characterized by their positions $\mathbf{r}$ and their axis $\hat{\mathbf {\nu}}=(\cos \theta, \sin \theta)$. The equations of motion for a particle $i$ are as follows
\begin{equation}
\begin{array}{lll}
\partial_t{\mathbf r}_i & = & v_0 \hat{\mathbf {\nu}_i} + \mu \sum_{j \ne i} {\mathbf F}_{ij} + {\mathbf \eta}^T_i (t)\\
\partial_t \theta_i & = & {\mathbf \eta}^R_i (t)\\
\end{array}
\end{equation}
in which $v_0$ denotes the self-propulsion velocity, $\mu$ the mobility, ${\mathbf \eta}_i^T (t)$ and $\eta^R_i (t)$ are the translational and rotational white noise terms, respectively. The force ${\mathbf F}_{ij}$ between two particles $i$ and $j$ is short-ranged and repulsive with, in this case, ${\mathbf F}_{ij} = - k (\frac{2a}{r_{ij}} - 1 ) \mathbf {r_{ij}}$ if $r_{ij} < 2a$ and $0$ otherwise, in which $k$ is a force constant and $a$ the radius of the particle. Other repulsive potentials have also been employed, such as {\it e.g.} a Weeks-Chandler-Andersen potential, leading to the same type of clustering and phase separation~\cite{redner2013structure}. Alternate models have also included attractive interactions between active particles~\cite{paliwal2018chemical,prymidis2016vapour,van2020predicting,prymidis2015self,navarro2015clustering,alarcon2017morphology}. We add that ABP exhibit similar features to those observed for Vicsek models, including the onset of giant fluctuations in the system~\cite{fily2012athermal}. The equations of motion can also be extended to account for active particles of different shapes~\cite{mallory2018active}, including dumbells~\cite{siebert2017phase,digregorio20192d,belan2021active,clopes2020hydrodynamic,tung2016micro,gonnella2014phase,petrelli2018active}, which opens the door to simulations of active nematics~\cite{keber2014topology,doostmohammadi2018active,pismen2013dynamics,dell2018growing,putzig2016instabilities,guillamat2018active,giomi2014spontaneous,mueller2019emergence,zhang2016dynamic,ellis2018curvature,rivas2020driven,giomi2012banding,shankar2018low,giomi2013defect,putzig2014phase,mishra2014aspects,decamp2015orientational,ramaswamy2003active,bertin2013mesoscopic}.\\

Cates and Tailleur~\cite{cates2015motility} proposed the concept of active simple fluids as fluids composed of spherical self-propelled particles, whose interactions are isotropic. Isotropic interactions encompass attractive and repulsive potential, as well as, {\it e.g.}, different types of chemical signaling and quorum sensing in bacteria~\cite{miller2001quorum}. These fluids were found to exhibit a far richer phase behavior than non-active, or passive, fluids which interact through the same isotropic potential. For instance, purely repulsive ABP undergo a liquid-gas phase separation~\cite{cates2013active,cates2015motility,caprini2020spontaneous,digregorio2018full,suma2014motility,solon2018generalized,partridge2019critical,van2019interrupted,patch2017kinetics,sese2018velocity}, while repulsive soft spheres do not. This phenomenon, termed as motility-induced phase separation (MIPS)~\cite{tailleur2008statistical,tjhung2018cluster}, arises from the nonequilibrium nature of the active fluid. It can be characterized as the coexistence of a dilute active gas with a dense liquid cluster of reduced motility.\\

\begin{figure}[ht]
\centering
\includegraphics[width=12cm]{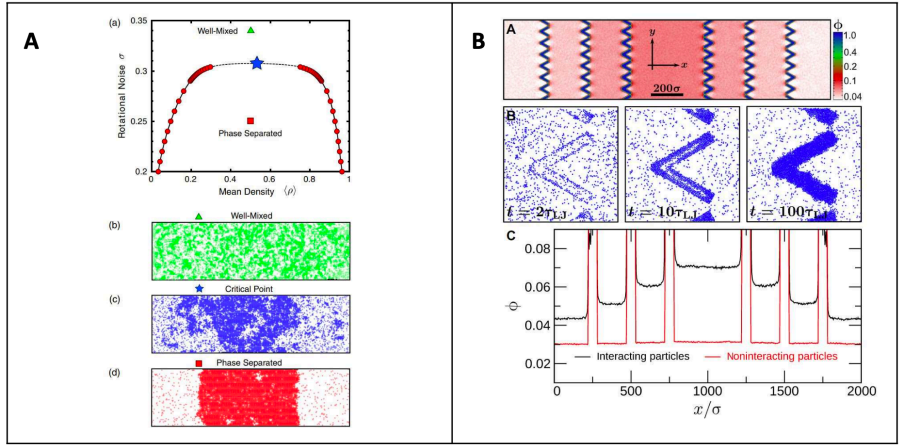}
\caption{A) Phase behavior of active particles, with (top panel) a phase diagram obtained from Monte Carlo simulations of a $72 \times 216$ lattice model. Snapshots shows in (second panel) a well-mixed region at high rotational noise ($\sigma$), (third panel) the system at the critical point and (bottom panel) a phase separated region at low noise~\cite{partridge2019critical}. Reprinted with permission from {\it Phys. Rev. Lett.}~ {\bf 2019}, 123, 068002. Copyright 2019 American Physical Society. B) Light-induced self-assembly of active rectification devices, with (top panel) the time-averaged packing fraction $\phi(r)$ in the steady-state for a system of self-assembled funnels, in (middle panel) a magnification of one of the funnels showing the self-assembly process at three different times starting from a homogeneous particle distribution, and in (bottom panel) the packing fraction $\phi$ as a function of $x$ and averaged over $y$ (in black) that shows the pumping effect (rectification) and the lack of rectification obtained for non-interacting particles~\cite{stenhammar2016light}. Reprinted with permission from {\it Sci. Adv.}~ {\bf 2016}, 2, e1501850. Copyright 2016  American Association for the Advancement of Science.}
\label{fig6}
\end{figure}

{\bf Thermodynamics for microswimmers}\\

Recent studies have focused on establishing the phase diagram (see Fig.~\ref{fig6}A) of ABPs and microswimmers suspensions~\cite{klamser2018thermodynamic,fily2014freezing,stenhammar2014phase,redner2013structure,wysocki2014cooperative,digregorio2018full,nie2020stability,omar2021phase,jeckel2019learning,be2020phase} and, more generally, to identify which thermodynamic concepts can account for the behavior of active matter~\cite{takatori2015towards}. For instance, it has been shown that, unlike in equilibrium systems, the pressure exerted by an active fluid on a wall does not always obey an equation of state but depends explicitly on the potential modeling the wall ~\cite{solon2015pressure,junot2017active}. Furthermore, while pressure in several dry active systems may obey an equation of state~\cite{di2010bacterial,yang2014aggregation,takatori2014swim}, it only partially fulfils the role played by pressure at equilibrium~\cite{mallory2014anomalous,sandford2017pressure,nikola2016active,speck2016ideal,marconi2016pressure,falasco2016mesoscopic,ginot2018sedimentation}. In such systems, pressure can serve to identify coexistence between phases but is unable to yield the correct binodals through a Maxwell construction~\cite{stenhammar2014phase}. This has led to the definition of the swim pressure~\cite{takatori2014swim,omar2020microscopic} that originates from the swimmers self-propulsion and reorientation and, as such, is an athermal quantity~\cite{takatori2016forces}. This apparent lack of an equation of state for pressure in active systems was traced back to the violation of momentum conservation in active systems. It was later found that an equation of state could be recovered provided that the self-propulsion dynamics were independent of the positional and angular degrees of freedom~\cite{fily2017mechanical}. This opened the door to the development of a generalized thermodynamics for active systems~\cite{rodenburg2017van,ginot2015nonequilibrium,solon2018generalized}. Recent work has focused on the concept of surface tension~\cite{paliwal2017non,bialke2015negative,zakine2020surface,wittmann2019pressure,li2021hierarchical,singh2020self,lauersdorf2021phase}, and shown that surface tension behaves differently for active systems from their equilibrium counterparts. In addition to an equation-of-state abiding contribution that constitutes a generalization of the Young-Laplace equation to active matter ~\cite{zakine2020surface}, the surface tension also includes a system-dependent contribution which incorporates information on the nature and interaction between the two materials at the junction.\\

Nonequilibrium and living systems are also associated with irreversibility and time-reversal symmetry breaking. In other words, the injection of energy into the system can give rise to the onset of a steady-state, the emergence of steady-state currents and of a global entropy production rate. This has been quantified by fluctuation theorems~\cite{gallavotti1995dynamical,evans1993probability,crooks1999entropy,jarzynski2000hamiltonian,evans2002fluctuation,van2003extension,maes1999fluctuation,kurchan1998fluctuation,seifert2005entropy,desgranges2020entropy,lebowitz1999gallavotti,hayashi2010fluctuation,ciliberto1998experimental,campisi2009fluctuation,gaspard2004fluctuation,lepri2000gallavotti}. The determination of the entropy production rate has drawn significant interest in recent years, and the development of several thermodynamics and data science-based approaches is currently under way~\cite{nardini2017entropy,razin2020entropy,borthne2020time,shankar2018hidden,mandal2017entropy,caballero2020stealth,pietzonka2017entropy,guo2021play}.\\

{\bf Dynamical behavior and external stimuli}\\

Trapping microswimmers to form static or dynamic patterns (see Fig.~\ref{fig6}B) is also an area of active research~\cite{bechinger2016active,stenhammar2016light}. One of the simplest ways to trap particles is to use spatial confinement. Fily {\it et al.} studied the dynamics of non-interacting and non-aligning self-propelled particles under strong confinement ({\it i.e.}, when the box dimension is smaller than the persistence length of the particles. Fily {\it et al.} found that they could modify the particles' spatial distribution by changing the geometry of the simulation box. For instance, particles are packed in areas of strong curvatures in a 2D ellipse-shaped container, while particles concentrate in sharp corners in a 2D polygon-shaped container. Moreover, the greater the persistence time for the particles, the longer particles will remain trapped~\cite{fily2014dynamics}. This could provide insight into how specific spatial distributions of bacteria arise in nature such as, {\it e.g.}, in biofilm formation. In systems of 2D active particles interacting via soft repulsive interactions, Yang {\it et al.} ~\cite{yang2014aggregation} found that particles aggregated on the sides of the container to form clusters, which left the center of the container virtually unoccupied. This form of segregation, which results from a combination of confinement and activity, could contribute to our understanding of cell sorting in embryonic development.

Both rigid and deformable microswimmers can exhibit intriguing trajectories and dynamical patterns in complex flows. Shape deformability, combined with the motility of microswimmers, add a layer of complexity. For instance, red blood cells can deform when they pass through microchannels that are smaller than the size of blood cells. Fedosov {\it et al.} employed mesoscale hydrodynamic simulations to predict the phase diagram for the shapes and dynamics of red blood cells in flow through cylindrical microchannels~\cite{fedosov2014deformation}. They found a rich dynamical behavior, with snaking and tumbling discocytes, slippers performing a swinging motion, and stationary parachutes. One of the simplest flows is a linear shear flow. Recently, Gaffney {\it et al.}~\cite{gaffney2022canonical} showed that shape-deforming swimmers moving in the plane of a shear flow followed Jeffery’s orbits~\cite{jeffery1922motion}. Tarama {\it et al.} proposed a 2D theoretical and numerical framework based on the dynamics of the particles' orientation and deformation to understand the behavior of deformable active particles under shear flow. They found a manifold of different dynamical modes, including active straight motion, periodic motions, motions on undulated cycloids, winding motions, as well as quasi-periodic and chaotic motions. The validity of the model was tested against experimental data on self-propelled droplets undergoing a linear shear flow.~\cite{tarama2013dynamics,tarama2017swinging} Numerous studies have focused on microswimmers in complex flows.~\cite{elgeti2015physics,chuphal2021effect}. For instance, Zottl and Stark~\cite{zottl2012nonlinear} studied the three-dimensional dynamics of a spherical microswimmer in cylindrical Poiseuille flow. They found that microswimmers display swinging and tumbling trajectories. In 2D, such trajectories are equivalent to oscillating and circling solutions of a pendulum. Hydrodynamic interactions between the swimmer and confining channel walls were found to lead to dissipative dynamics and result in stable trajectories, different for pullers and pushers. Most biological fluids such as mucus and blood are viscoelastic and non-Newtonian. However, most simulation studies on motile microorganisms have focused on Newtonian fluids so far. Recently, Mathijssen {\it et al.}  developed a model for microswimmer in non-Newtonian Poiseuille flows. Unlike Newtonian fluids, swimmers’ trajectories show oscillatory motion about the centerline. More specifically, swimmers in shear-thickening (-thinning) fluids migrate upstream more (less) quickly than in Newtonian fluids. The direct upstream migration is related to viscoelastic normal stress differences~\cite{mathijssen2016upstream}. We finally that the determination of transport coefficients of active fluids is an active area of research~\cite{banerjee2017odd,slomka2017geometry,giomi2010sheared,hargus2020time}.\\

\section{Conclusions and Perspectives}

In this review, we discuss recent advances in the design, synthesis and modeling of active fluids at the nano- and micro-scale. The design of active materials on both scales is driven by observations of real-life, biological, active systems. On the nanoscale, these include biological nanomachines, nanomotors that power molecular machines and enzymes as energy transducers. On the microscale, examples of biological systems belong to the microbial world with bacteria and algae, among others. Advances in synthetic methods have led, on the nanoscale, to the development of an artificial biology, artificial enzymes or nanozymes, fuel-dependent or fuel-free nanomotors and to the directed motion of nanomotors for therapeutics. On the microscale, synthetic active fluids encompass suspensions of active colloids and Janus micromotors, the design of modular microswimmers and their directed self-assembly, the control of microswimmers in complex environment and of collective motion in active fluids to achieve the synthesis of living materials. Theoretical advances include the development of a continuum theory for phoretic propulsion, of coarse-grained simulation methods for nanomotors, the simulation-aided design of nanomotors and control of their motion. On the microscale, progress has been made through the proposal of minimal models that capture the essential features of active matter, from the well-known Vicsek model for dry active matter to the Active Brownian Particle model for wet active matter. The latter has provided novel insight into the behavior of active matter, including the onset of a motility-induced phase separation as a general feature of active fluids.

Active fluids are a fascinating, as well as thriving research field. Ongoing research focuses on the design of multi-component, modular, programmable and adaptative nano- and micro-machines. On the theoretical front, current developments focus on the establishment of a thermodynamics for microswimmers, measures for entropy production in these far-from-equilibrium systems and the understanding of their dynamical behavior and collective response to external stimuli. It is expected that concerted experimental and theoretical efforts will lead to the emergence of an autonomous soft matter robotics, able to perform tasks in a controlled manner on both scales. We conclude this review by highlighting two rapidly emerging areas of research on active matter. First, the application of machine learning and artificial intelligence to these systems promises to shed light on the behavior of these systems and to provide new ways to control their function~\cite{cichos2020machine}. Recent work has shown how phase changes, and the inset of MIPS, can be detected via the use of fully connected networks in conjunction with graph neural networks~\cite{dulaney2021machine}. Another example is the application of a convolutional long-short-term-memory (ConvLSTM) algorithm for the forecasting of the dynamics of active nematics~\cite{zhou2021machine}. Combining autoencoders and recurrent neural networks with residual architecture has also recently allowed to map out the spatiotemporal variation
of multiple hydrodynamic parameters and forecast the chaotic
dynamics of active nematics solely from
image sequences of their past~\cite{colen2021machine}. Most notably, recent work has shown how the adaptative behavior and learning, usually demonstrated by living systems, could be extended to synthetic active particles via reinforcement learning~\cite{muinos2021reinforcement,falk2021learning,colabrese2017flow,gerhard2021hunting}. The second emerging area of research focuses on the use of active matter to design and build soft matter robots. For instance, recent work has shown how the control of self-propelled colloidal particle propulsion speeds can lead to the cooperative capture and transport of cargo particles~\cite{yang2020cargo}. Very interestingly, such control can be exerted through the application of specific light patterns both in the case of synthetic active particles~\cite{stenhammar2016light,schmidt2019light,lozano2016phototaxis}, but also in the case of bacteria~\cite{arlt2018painting,arque2019intrinsic,frangipane2018dynamic}. Indeed, {\it E. coli} cells under anaerobic conditions can express proteorhodopsin, a green-photon-driven proton pump~\cite{beja2000bacterial}, and have their self-propulsion velocity controlled via green light illumination. Many other types of devices leveraging active matter are currently under development, including the applications of active soft materials, including actin-, tubulin-, and cell-based systems, to perform logic operations and thus perform computations~\cite{zhang2022logic}.\\

\begin{acknowledgments}
This material is based upon work supported by the U.S. Department of Energy, Office of Science, Office of Basic Energy Sciences under Award Number DE-SC-0020976.\\
\end{acknowledgments}

{\bf AUTHOR DECLARATIONS}\\

{\bf Conflict of interest}\\

The authors have no conflict to disclose.\\

{\bf Data availability statement}\\

Data sharing is not applicable to this article as no new data were created or analyzed in this study.\\

\bibliography{Design}

\end{document}